\documentclass{aa}  
\usepackage{graphics}
\usepackage{graphicx}
\usepackage{multicol}
\usepackage[varg]{txfonts}

\usepackage{natbib,twoopt}
\usepackage{amsmath} 
\usepackage[breaklinks=true]{hyperref} 
\bibpunct{(}{)}{;}{a}{}{,} 
\makeatletter
\newcommandtwoopt{\citeads}[3][][]{\href{http://adsabs.harvard.edu/abs/#3}%
{\def\hyper@linkstart##1##2{}%
\let\hyper@linkend\@empty\citealp[#1][#2]{#3}}}
\newcommandtwoopt{\citepads}[3][][]{\href{http://adsabs.harvard.edu/abs/#3}%
{\def\hyper@linkstart##1##2{}%
\let\hyper@linkend\@empty\citep[#1][#2]{#3}}}
\newcommandtwoopt{\citetads}[3][][]{\href{http://adsabs.harvard.edu/abs/#3}%
{\def\hyper@linkstart##1##2{}%
\let\hyper@linkend\@empty\citet[#1][#2]{#3}}}
\newcommandtwoopt{\citeyearads}[3][][]%
{\href{http://adsabs.harvard.edu/abs/#3}
{\def\hyper@linkstart##1##2{}%
\let\hyper@linkend\@empty\citeyear[#1][#2]{#3}}}
\makeatother

\usepackage{color}
\definecolor{mygreen}{RGB}{0,128,0}
\definecolor{mygray}{gray}{0.6}

\newcommand{\GMnom} {\hbox{($\mathcal{GM})^{\rm N}_\odot$}}
\newcommand{\Rnom} {\hbox{$\mathcal{R}^{\rm N}_\odot$}}
\newcommand{\Lnom} {\hbox{$\mathcal{L}^{\rm N}_\odot$}}
\hypersetup{colorlinks=true,linkcolor=blue,citecolor=blue,urlcolor=blue}

\def\wmean{1.00481}
\def\werr{0.00412}
\def\wrmspercent{0.41}

\def\meanwave{1.6513}
\def\meanwaveerr{0.0067}
\def\diffmagAB{0.535}
\def\diffmagABerr{0.016}
\def\thetaLDA{8.502} 
\def\thetaLDAerrtot{0.038} 
\def\thetaLDArelerr{0.43} 
\def\alphaA{0.1404}
\def\alphaAerr{0.0050}
\def\alphaArelerr{3.6} 
\def\thetaLDB{5.999} 
\def\thetaLDBerrtot{0.025} 
\def\thetaLDBrelerr{0.42} 
\def\alphaB{0.1545}
\def\alphaBerr{0.0044}
\def\alphaBrelerr{2.8} 
\def\alphasun{0.15027}

\def\radiusA{1.2234} 
\def\radiusAerrstat{0.0013} 
\def\radiusAerrsyst{0.0051} 
\def\radiusAerrtot{0.0053} 
\def\radiusArelerr{0.43}  

\def\radiusAm{8.511} 
\def\radiusAerrm{0.037} 
\def\radiusB{0.8632} 
\def\radiusBerrstat{0.0009} 
\def\radiusBerrsyst{0.0036} 
\def\radiusBerrtot{0.0037} 
\def\radiusBrelerr{0.43} 

\def\radiusBm{6.006} 
\def\radiusBerrm{0.026} 
\def\ratioR{1.4172} %
\def\ratioRerr{0.0016} %
\def\ratioRrelerr{0.11} 
\def\massA{1.1055}
\def\massAerr{0.0039}
\def\massAkg{2.1983}
\def\massAerrkg{0.0078}
\def\massB{0.9373}
\def\massBerr{0.0033}
\def\massBkg{1.8638}
\def\massBerrkg{0.0066}
\def\lumA{1.521}
\def\lumAerr{0.015}
\def\lumAW{5.821} 
\def\lumAerrW{0.058}
\def\lumB{0.503}
\def\lumBerr{0.007}
\def\lumBW{1.925} 
\def\lumBerrW{0.026}
\def\teffA{5795}
\def\teffAerr{19}
\def\teffB{5231}
\def\teffBerr{21}
\def\loggA{4.3117}
\def\loggAerr{0.0015}
\def\loggB{4.5431}
\def\loggBerr{0.0015}

\begin{document} 


\title{The radii and limb darkenings of $\alpha$~Centauri A and B\thanks{The interferometric data are available in OIFITS format via anonymous ftp to \url{http://cdsarc.u-strasbg.fr} (\url{http://130.79.128.5}) or via \url{http://cdsweb.u-strasbg.fr/cgi-bin/qcat?J/A+A/}}
}
\subtitle{Interferometric measurements with VLTI/PIONIER}
\titlerunning{The radii and limb darkenings of $\alpha$\,Cen A \& B}
\authorrunning{P. Kervella et al.}

\author{
P.~Kervella\inst{1,2}
\and
L.~Bigot\inst{3}
\and
A.~Gallenne\inst{4}
\and
F.~Th\'evenin\inst{3}
}
\institute{
Unidad Mixta Internacional Franco-Chilena de Astronom\'{i}a (CNRS UMI 3386), Departamento de Astronom\'{i}a, Universidad de Chile, Camino El Observatorio 1515, Las Condes, Santiago, Chile, \email{pkervell@das.uchile.cl}.
\and
LESIA (UMR 8109), Observatoire de Paris, PSL Research University, CNRS, UPMC, Univ. Paris-Diderot, 5 Place Jules Janssen, 92195 Meudon, France, \email{pierre.kervella@obspm.fr}.
\and
Universit\'e C\^ote d'Azur, Observatoire de la C\^ote d'Azur, CNRS, Lagrange UMR 7293, CS 34229, 06304, Nice Cedex 4, France.
\and
European Southern Observatory, Alonso de C\'ordova 3107, Casilla 19001, Santiago 19, Chile.
}
\date{Received ; Accepted}
 
\abstract
{The photospheric radius is one of the fundamental parameters governing the radiative equilibrium of a star.
We report new observations of the nearest solar-type stars $\alpha$\,Centauri A (G2V) and B (K1V) with the VLTI/PIONIER optical interferometer.
The combination of four configurations of the VLTI enable us
to measure simultaneously the limb darkened angular diameter $\theta_\mathrm{LD}$ and the limb darkening parameters of the two solar-type stars in the near-infrared $H$ band ($\lambda = 1.65\,\mu$m).
We obtain photospheric angular diameters of $\theta_\mathrm{LD}(A) = \thetaLDA \pm \thetaLDAerrtot$\,mas ($\thetaLDArelerr\%$) and $\theta_\mathrm{LD}(B) = \thetaLDB \pm \thetaLDBerrtot$\,mas ($\thetaLDBrelerr\%$), through the adjustment of a power law limb darkening model.
We find $H$ band power law exponents of $\alpha(A) = \alphaA \pm \alphaAerr$ ($\alphaArelerr\%$) and $\alpha(B) = \alphaB \pm \alphaBerr$ ($\alphaBrelerr\%$), which closely bracket the observed solar value ($\alpha_\odot = \alphasun$).
Combined with the parallax $\pi = 747.17 \pm 0.61$\,mas previously determined, we derive linear radii of $R_A = \radiusA \pm \radiusAerrtot\ R_\odot$ ($\radiusArelerr\%$) and $R_B = \radiusB \pm \radiusBerrtot\ R_\odot$ ($\radiusBrelerr\%$).
The power law exponents that we derive for the two stars indicate a significantly weaker limb darkening than predicted by both 1D and 3D stellar atmosphere models.
As this discrepancy is also observed on near-infrared limb darkening profile of the Sun, an improvement of the calibration of stellar atmosphere models is clearly needed.
The reported PIONIER visibility measurements of $\alpha$\,Cen~A and B provide a robust basis to validate the future evolutions of these models.
}

   \maketitle
%

\section{Introduction}

The photospheric radius of a star is intimately linked to its radiative equilibrium through the Stefan-Boltzmann law.
Over the lifetime of a star, the radius and effective temperature are directly related to the production of nuclear energy at the core of the star.
Measured radii and seismic oscillation frequencies provide complementary constraints for stellar structure and evolution models (\citeads{2007ApJ...659..616C}; \citeads{2007A&ARv..14..217C}; \citeads{2015ApJ...811L..37M}).
The triple system $\alpha$ Centauri (\object{WDS J14396-6050AB}, \object{GJ559AB}) is our closest stellar neighbor, at a distance of only $d = 1.3384 \pm 0.0011$\,pc for the main components A and B ($\pi = 747.17 \pm 0.61$\,mas; \citeads{2016kervella}).
The third member is \object{Proxima} (M5.5V, \object{GJ551}).
$\alpha$\,Cen A and B are dwarf stars of spectral types G2V (A, \object{HD 128620}) and K1V (B, \object{HD 128621}).
%
%
Their proximity and similarity to the Sun in terms of mass and spectral type make $\alpha$\,Cen extremely attractive from the standpoint of stellar physics (\citeads{2016MNRAS.460.1254B}; \citeads{2016A&A...586A..90P}; \citeads{2016JPhCS.665a2081B}; \citeads{2015A&A...577A..60S}; \citeads{2015AJ....149...58A}; \citeads{2015A&A...573L...4L}; \citeads{2014A&A...566A..98B}) and extrasolar planet research (\citeads{2016AJ....151..111Q}; \citeads{2016MNRAS.456L...6R}; \citeads{2015MNRAS.450.2043D}; \citeads{2016arXiv160703090W}; \citeads{2014MNRAS.444.2167A}; \citeads{2013ApJ...777..165K}; \citetads{2012Natur.491..207D}).
In addition, the $\alpha$\,Cen pair is one of the principal benchmark stars of the \emph{Gaia} mission (\citeads{2015A&A...582A..49H}; \citeads{2015A&A...582A..81J}).
An extremely accurate calibration of its fundamental parameters is essential for the validation of the data analysis methods that are currently applied to the fainter targets of the \emph{Gaia} catalog (see e.g.~\citeads{2013A&A...559A..74B}).

We present new optical interferometric measurements of $\alpha$\,Cen~A and B obtained with the VLTI/PIONIER interferometer in the near-infrared $H$ band.
These measurements are intended to update and improve those reported more than a decade ago by \citetads{2003A&A...404.1087K} and \citetads{2006A&A...446..635B}.
The details of these observations are described in Sect.~\ref{observations}, where we also report the observation of a binary star (HD\,123999) that we employ to calibrate the effective wavelength of the instrument.
Sect.~\ref{analysis} is dedicated to the derivation of the angular diameters of the two stars together with their limb darkening parameters.
We also check for the possible presence of spots on the surfaces of the two stars or close-in faint companions.
We compare in Sect.~\ref{discussion} the derived limb darkening parameters with existing models from the literature or from new 3D hydrodynamical simulations.

\section{Observations and data reduction\label{observations}}

\subsection{$\alpha$\,Cen observations}

\begin{table}
	\caption{Calibrators of $\alpha$\,Cen and HD\,123999.
	They were selected from the \citetads{2005A&A...433.1155M} (M05) and \citetads{2010yCat.2300....0L} (L10) catalogs.}
	\label{calibrators}
	\centering
	\renewcommand{\arraystretch}{1.2}
	\begin{tabular}{lcccccc}
		\hline\hline
		HD	& $\rho$\,$^a$ & Spect. & $m_{V}$ & $m_{H}$ & $\theta_\mathrm{UD}$\,$^b$ & Cat.\\ 
		\hline		\noalign{\smallskip}
	        \multicolumn{7}{c}{$\alpha$\,Cen A \& B} \\
		\object{127753} & 4.3 & K5III & 7.08 & 2.74 &  $1.683_{0.023}$ & M05 \\  
		\object{133869} & 3.7 & K3III & 8.02 & 3.72 &  $1.043_{0.015}$ & M05 \\
		\hline		\noalign{\smallskip}
	        \multicolumn{7}{c}{HD\,123999} \\
		119126 & 7.2 & G9III & 5.63 & 3.59 & $0.974_{0.013}$ & M05 \\
		123612 & 0.9 & K5III &  6.55  & 3.28 & $1.268_{0.091}$ & L10 \\
		125728 & 2.5 & G8II & 6.79 & 4.61 & $0.560_{0.040}$ & L10 \\
		128402 & 6.1 & K0 & 6.38 & 4.22 & $0.762_{0.054}$ & L10 \\
		130948 & 9.1 & F9IV & 5.88 & 4.69 & $0.557_{0.039}$ & L10 \\
		\hline
	\end{tabular}
        \tablefoot{
        $^a$ $\rho$ is the angular separation from $\alpha$\,Cen or HD\,123999 in degrees.
        $^b$ $\theta_\mathrm{UD}$ is the uniform disk angular diameter in the $H$ band, in milliarcseconds.
        }
\end{table}

We observed $\alpha$\,Cen A and B using the Very Large Telescope Interferometer \citepads{2014SPIE.9146E..0JM} equipped with the PIONIER beam combiner (\citeads{2010SPIE.7734E..99B}, \citeads{Le-Bouquin:2011fj}) operating in the infrared $H$ band ($\lambda = 1.6\,\mu$m).
The stellar light was collected by the four 1.8\,m Auxiliary Telescopes.
We used four configurations of the interferometer\footnote{\url{https://www.eso.org/sci/facilities/paranal/telescopes/vlti.html}}: A0-B2-C1-D0, D0-K0-G2-J3, A0-J2-G1-J3 and the transition quadruplet A0-G2-D0-J3.
Thanks to the Earth rotation supersynthesis, these four configurations provided an almost continuous coverage of the interferometric visibility function of the two stars up to the third lobe of component A and the second lobe of component B.
This means that the apparent disk of the two stars were resolved with two to three resolution elements of the interferometer, thus enabling the measurement of their limb darkening (hereafter LD) in addition to their angular diameters.
Such a complete coverage of the visibility function of dwarf stars is, to our knowledge, unprecedented.
The observations were obtained on four nights: 23, 27, 29, and 30 May 2016, under good seeing conditions (visible DIMM seeing between 0.6 and $1.1\arcsec$).
The pointings of $\alpha$\,Cen were interspersed with observations of calibrator stars (Table~\ref{calibrators}) to estimate the interferometric transfer function of the instrument.
These stars were selected close angularly to $\alpha$\,Cen to avoid a polarimetric mismatch of the beams.
The raw data have been processed using the standard {\tt pndrs}\footnote{\url{http://www.jmmc.fr/data_processing_pionier.htm}} data reduction software \citepads{Le-Bouquin:2011fj}.

\subsection{Wavelength calibration\label{wavecalibration}}

\subsubsection{Importance of wavelength calibration}

The equivalent of the plate scale of an interferometer is the instrumental angular resolution, which is defined as the ratio of the effective observation wavelength $\lambda$ and the projected baseline length $B$.
Any dimensional measurement  derived from interferometric observations (e.g., angular diameters and binary star separation) is directly proportional to the instrumental angular resolution, and its associated systematic uncertainty directly translates to the measured quantities.
It should be noted that differential measurements (e.g., angular diameter ratios and LD coefficients) are not affected by this systematic uncertainty.
The internal (laboratory) calibration of the PIONIER wavelength scale has an accuracy of about 1\% \citepads{Le-Bouquin:2011fj}, which is a limiting factor for the determination of the angular diameters of $\alpha$\,Cen A and B.
The projected baseline length $B$ is known to better than 1\,cm, i.e.~to a relative accuracy $< 10^{-4}$.
This is demonstrated by the fact that we consistently find the interference fringes within at most a few millimeters from the expected optical path difference.
As the angular diameter scales linearly with the baseline length, the baseline uncertainty is therefore negligible in the total error budget compared to the wavelength uncertainty.

\subsubsection{HD\,123999 as dimensional calibrator}

To improve the calibration of the wavelength scale of PIONIER, we adopt \object{HD 123999} (\object{12 Boo}, \object{HR 5304}) as a dimensional calibrator.
This original approach takes advantage of the fact that HD\,123999 has been the target of intense spectroscopic and interferometric observations for at least two decades (\citeads{2000ApJ...536..880B}; \citeads{2005ApJ...627..464B}; \citeads{2006AJ....131.2652T}; \citeads{2010ApJ...719.1293K}; \citeads{2011AJ....142....6B}).
As a result, its orbital parameters are known with an exquisite accuracy.
The two components have almost equal masses close to $1.4\,M_\odot$ and effective temperatures around 6150\,K (\citeads{2007MNRAS.377..373M}; \citeads{2010ApJ...719.1293K}). The combined spectral type of the binary is F8V \citepads{1538-3881-121-4-2148}, which is not far from $\alpha$\,Cen~A (G2V).
With an $H$-band apparent magnitude of $m_H = 3.6$, it is easily observable with PIONIER at high signal-to-noise ratio.

We obtained a series of ten observations of HD\,123999 between 21 February and 30 May 2016, interspersed with calibrator stars (Table~\ref{calibrators}).
We adjusted a classical binary model to the PIONIER squared visibilities, phase closure, and triple amplitude using the {\tt CANDID} tool\footnote{\url{https://github.com/amerand/CANDID}} \citepads{2015A&A...579A..68G}.
We checked that the {\tt LITpro} tool\footnote{\url{http://www.jmmc.fr/litpro}} \citepads{2008SPIE.7013E..44T} gives statistically compatible results. 
According to \citetads{2005ApJ...627..464B}, the angular diameters of the two components of HD\,123999 are $\theta_\mathrm{LD}(A) = 0.64$\,mas and $\theta_\mathrm{LD}(B) = 0.48$\,mas. They are slightly resolved by the interferometer, but cannot be accurately derived from our data. We therefore fix these parameters in the binary model fit.
The measured positions of HD\,123999\,B relative to A are listed in Table~\ref{12boo-positions}.

\subsubsection{Orbital solution}

\begin{figure*}[]
        \centering
        \includegraphics[width=16cm]{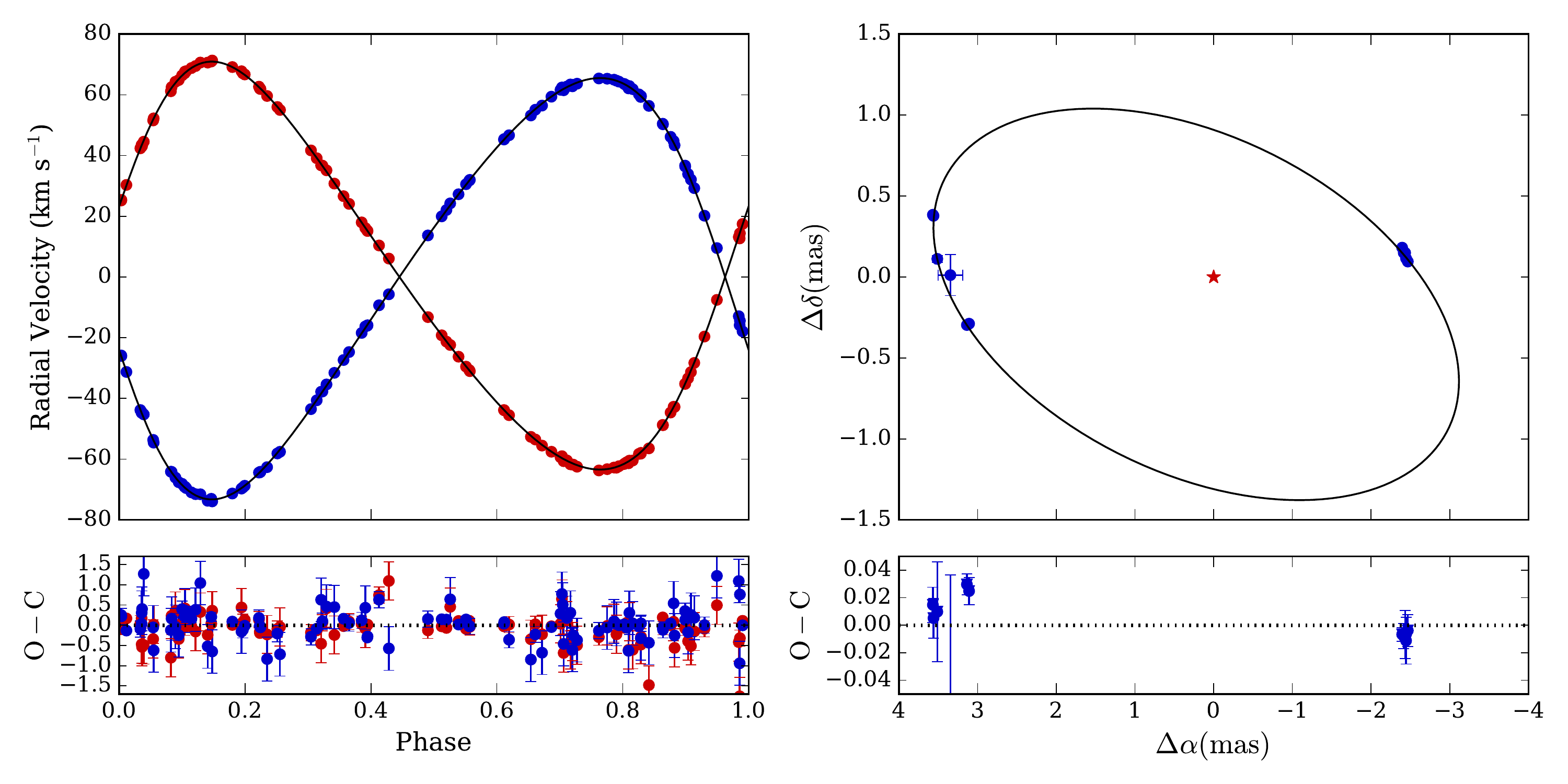}
        \caption{Radial velocities (left) and PIONIER astrometry (right) of HD\,123999. The quantities related to components A and B are represented using red and blue symbols, respectively, and are shown over the best-fit orbit.
        \label{orbit-12Boo}}
\end{figure*}

We considered for the fit of the orbital parameters the same radial velocity measurements as \citetads{2005ApJ...627..464B}, \citetads{2006AJ....131.2652T}, and \citetads{2010ApJ...719.1293K}, together with our new PIONIER astrometry.
The date of each measurement was converted to Heliocentric Julian Date (HJD) values.
The best-fit orbital parameters are listed in Table~\ref{parametersHD123999}.
The listed uncertainties are purely statistical and therefore do not contain the wavelength calibration uncertainty.
It is generally estimated to $0.5\%$ for PTI, as for example by \citetads{2005ApJ...627..464B}, but was not taken properly into account in the determination of the orbital parameter uncertainties by \citetads{2010ApJ...719.1293K}, as they state a relative uncertainty of only 0.16\% for the semi-major axis $a$.
The measured radial velocities and astrometry are shown over the adjusted orbit in Fig.~\ref{orbit-12Boo}.
All the parameter values that we obtain are consistent with previous works, except for the position angle of the ascending node $\Omega$, which is different by $180^\circ$.
The PIONIER measurements provide a significant improvement over previous interferometric measurements by the Palomar Testbed Interferometer (PTI; used by \citeads{2005ApJ...627..464B} and \citeads{2010ApJ...719.1293K}). For each epoch PIONIER produces simultaneously six squared visibilities and four independent closure phases, while PTI produces only a single squared visibility.
The PIONIER data thus provide a comprehensive data set to constrain the orbit orientation on sky very efficiently.
\citetads{2010ApJ...719.1293K} and \citetads{2000ApJ...536..880B} arbitrarily selected the $\Omega$ value smaller than $180^\circ$, as there is a $180^\circ$ ambiguity on $\Omega$ for two-telescope $V^2$ measurements.
\citetads{2005ApJ...627..464B} used NPOI data that include closure phases, and mentioned that they were able to constrain $\Omega$.
They obtained a projected orbit that is different from \citetads{2000ApJ...536..880B}, but in agreement with ours, i.e. with $\Omega \approx 260^\circ$.
However, for an unclear reason, they list a value of $\Omega < 180^\circ$ in their Table~4.
We also derive an average photometric contrast between the two components of HD\,123999 in the $H$ band of $\Delta m_H = \diffmagAB \pm \diffmagABerr$, which is in good agreement with \citetads{2005ApJ...627..464B} ($\Delta m_H = 0.56 \pm 0.02$) but not with \citetads{2010ApJ...719.1293K} ($\Delta m_H = 0.66 \pm 0.03$).

\begin{table}
        \caption{Orbital parameters determined for HD\,123999 from PIONIER observations and radial velocity measurements (the listed error bars include only statistical errors and should not be considered accurate at this level).}
        \centering          
        \label{parametersHD123999}
        \begin{tabular}{llll}
	\hline\hline
        \noalign{\smallskip}
        Parameter & Value \\
         \hline
        \noalign{\smallskip}
Period $P$ & $9.604559 \pm 0.000005$\,days \\
HJD of periastron $T_p$ & $2454100.4340 \pm 0.0006$ \\
Excentricity $e$ & $0.1924 \pm 0.0001$ \\
Long.~of periastron $\omega$ & $286.86 \pm 0.02\,\deg$ \\
Ascending node $\Omega$ & $260.71 \pm 0.12\,\deg$ \\
Semi-amplitude $K_A$ & $67.179 \pm 0.006$\,km\,s$^{-1}$ \\
Semi-amplitude $K_B$ & $69.390 \pm 0.002$\,km\,s$^{-1}$ \\
System velocity $\gamma$ & $9.62 \pm 0.03$\,km\,s$^{-1}$  \\
Semi-major axis $a$ & $3.436 \pm 0.003$\,mas \\
Inclination $i$ & $108.7 \pm 0.2\,\deg$ \\
$\chi^2_\mathrm{red}$ (deg. of freedom = 222) & 4.62 \\
Mass of A component $m_A$ & $1.432 \pm 0.005\,M_\odot$ \\
Mass of B component $m_B$ & $1.387 \pm 0.004\,M_\odot$ \\
Distance $d$ & $36.36 \pm 0.06$\,pc \\
Semi-major axis $a$ & $0.1249 \pm 0.0001$\,AU \\
Residual $V_A$ & $\sigma = 0.370$\,km\,s$^{-1}$ \\
Residual $V_B$ & $\sigma = 0.407$\,km\,s$^{-1}$ \\
   	\hline
        \end{tabular}
\end{table}

\subsubsection{Wavelength scaling factor\label{scalingfactor}}

We determine the scaling factor to be applied to the effective wavelength scale of PIONIER by comparing our best-fit angular semi-major axis $a_\mathrm{K16}=3.436$\,mas and the average of the values determined by us, \citetads{2005ApJ...627..464B}, $a_\mathrm{B05}=3.451$\,mas and \citetads{2010ApJ...719.1293K}, $a_\mathrm{K10} = 3.4706$\,mas:
\begin{equation}\label{eq:gamma}
\gamma = \frac{a_\mathrm{B05} + a_\mathrm{K10} + a_\mathrm{K16}}{3\ a_\mathrm{K16}} = \wmean \pm \werr
\end{equation}
%
%
We adopt as uncertainty for this quantity the fractional standard deviation of the three measurements of $a$.
This is a conservative approach, but it ensures that systematic differences of the wavelength scale calibration are properly taken into account in the error bars.
We thus multiplied the PIONIER wavelength scale by $\gamma = \wmean$ for the subsequent analysis, and quadratically added an uncertainty of $\wrmspercent\%$ to all derived angular diameters.
Including our value $a_\mathrm{K16}$ in the numerator of $\gamma$ in Eq.~\ref{eq:gamma} induces a degree of circularity in the resulting calibration.
But we prefer this approach as the (unweighted) averaging of the three measurements of $a$ results in principle in a lower potential bias on the wavelength calibration.
If we simply consider the ratio of the average of $a_\mathrm{B05}$ and $a_\mathrm{K10}$ to our measurement ($a_\mathrm{K16}$) we obtain $\gamma = 1.00722$.
This is only $+0.6\,\sigma$ away from the value we adopt, and both approaches are thus statistically equivalent.
The six spectral channels have central wavelengths of $\lambda = 1.52, 1.57,  1.62, 1.67, 1.72,$ and $1.77\,\mu$m with a bandwidth of 48\,nm per channel.
The mean effective observing wavelength of PIONIER is $\lambda_0 = \meanwave \pm \meanwaveerr\,\mu$m.

In addition to HD\,123999, we obtained two PIONIER measurements of the binary \object{HD 78418}, whose orbital parameters were also determined by \citetads{2010ApJ...719.1293K}.
Although these two points are insufficient to conduct the same extensive analysis as with HD\,123999, we derive a value of
$\gamma = 1.00169$ from the comparison of the observed and predicted separations.
This is in good agreement ($-0.8\sigma$) with the $\gamma$ factor determined from the observations of HD\,123999.

The distance we obtain for HD\,123999 after scaling by the $\gamma$ factor ($d = 36.18 \pm 0.15$\,pc) is consistent with the original \emph{Hipparcos} value ($d = 36.7 \pm 1.0$\,pc, \citeads{1997ESASP1200.....E}).
This is an indirect indication that our wavelength calibration is not biased by a large amount.
The revised processing of the \emph{Hipparcos} data by \citetads{2007A&A...474..653V} gives a distance of $d = 37.4 \pm 0.3$\,pc, that is $+3.6\,\sigma$ from our value.
However, HD\,123999 is listed as a single object in both the original and revised catalogs.
The flux ratio $f_A/f_B \approx 1.6$ of the two components (in the $H$ band) is significantly larger than their mass ratio $m_A/m_B \approx 1.03$.
This induces a wobble of the center of light during the orbital cycle of the pair, which has a range of approximately $\pm 0.9$\,mas ($\pm 3\%$ of the parallax value).
The short orbital period of the system averaged out this wobble over the series of \emph{Hipparcos} observations that were used to compute the parallax, but the error bar of the revised parallax by \citetads{2007A&A...474..653V} may be underestimated.
For this reason, we prefer not to rely on the \emph{Hipparcos} distance to HD\,123999 to calibrate the wavelength of PIONIER.
The coming \emph{Gaia} distance to the system will eventually provide us with a much more accurate calibration of the distance.
This will reduce our wavelength calibration systematic uncertainty to $0.1\%$.
Also, the new GRAVITY instrument of the VLTI (\citeads{2011Msngr.143...16E}; \citeads{2014A&A...567A..75L}) within the next few years will provide interferometric visibilities and closure phases with a wavelength accuracy in the $K$ band ($\lambda = 2.2\,\mu$m) better than $10^{-4}$ owing to a dedicated internal reference laser source.
%

\section{Analysis\label{analysis}}

\subsection{Limb darkened disk angular diameters\label{LDmodelfit}}

\subsubsection{Parametric limb darkening models\label{parametric}}

For a reasonably good constraint of an N-parameter LD model, it is necessary to sample properly the maximum of the lobe with order N+1 of the visibility function.
For this reason, we cannot adjust the classical four-parameter non-linear law that is usually presented in the literature as an approximation to the intensity profiles derived from atmosphere models (see e.g.~\citeads{2011A&A...529A..75C}).
We however include it here with fixed coefficients taken from the literature for comparison.
As we measure the visibility function of $\alpha$\,Cen~A up to the beginning of the fourth lobe, a two-parameter model is the maximum acceptable.
For $\alpha$\,Cen~B, we measure the maximum of the second lobe and part of the third lobe, so we can only fit a single-parameter model.
Different types of one- and two-parameter LD models are classically employed to approximate the results from the stellar atmosphere models:
\begin{itemize}
\item uniform disk:
	\begin{equation} \label{udmodel}
	I(\mu)/I(1) = 1
	\end{equation}
\item linear:
	\begin{equation} \label{linmodel}
	I(\mu)/I(1) = 1-u\,(1-\mu)
	\end{equation}
\item power law \citepads{1997A&A...327..199H}:
	\begin{equation} \label{powmodel}
	I(\mu)/I(1) = \mu^\alpha
	\end{equation}
\item quadratic:
	\begin{equation} \label{quadmodel}
	I(\mu)/I(1) = 1-a\,(1-\mu)-b\,(1-\mu)^2
	\end{equation}
\item square root:
	\begin{equation} \label{sqrtmodel}
	I(\mu)/I(1) = 1 - c\,(1 - \mu) - d\,(1 - \sqrt{\mu})
	\end{equation}
\item four-parameter:
	\begin{equation}
	I(\mu)/I(1) = 1 - \sum_{k=1}^{4}a_k \left(1-\mu^{k/2}\right)
	\end{equation}
\end{itemize}
%
In addition, we consider the following polynomial model with six parameters:
\begin{itemize}
\item polynomial:
	\begin{equation} \label{solarld}
	I(\mu)/I(1) = \frac{\sum_{k=0}^{5}s_k\,\mu^k}{\sum_{k=0}^{5}s_k}
	\end{equation}
\end{itemize}
This type of model cannot be adjusted directly to the PIONIER data because there are too many parameters.
However, it is particularly interesting as it reproduces very well the observed intensity profile of the Sun measured by \citetads{1977SoPh...52..179P} in the infrared.
We computed the average parameters for the solar profile over the wavelength range covered by PIONIER using the parameters listed by these authors, and we obtain:
	\begin{align} \label{solarmodel}
	I(\mu) =\  & 0.59045 + 1.41938\,\mu -3.01866\,\mu^2 + \\
	 & 3.99843\,\mu^3 -2.67727\,\mu^4 + 0.68758\,\mu^5 \nonumber
	\end{align}
To be able to adjust a realistic LD model to the PIONIER squared visibilities, we consider the $s_0$ coefficient (constant term of the polynomial expression) as a variable parameter, while keeping the $s_1$ to $s_5$ parameters to their solar values (Eq.~\ref{solarmodel}).
The division of $I(\mu)$ by $I(1) = \sum_{k=0}^{5}s_k$ in the fit of the model results in an overall scaling of the polynomial coefficients of the normalized profile.
The adoption of this scaled solar LD profile is physically justified by the fact that the spectral types of both components of $\alpha$\,Cen are close to solar (in particular A).
This innovative approach allows us to preserve the shape of the observed LD profile of the Sun, while scaling it using a single parameter to match the measured PIONIER visibilities of $\alpha$\,Cen A and B.
\begin{figure}[]
        \centering
        \includegraphics[width=\hsize]{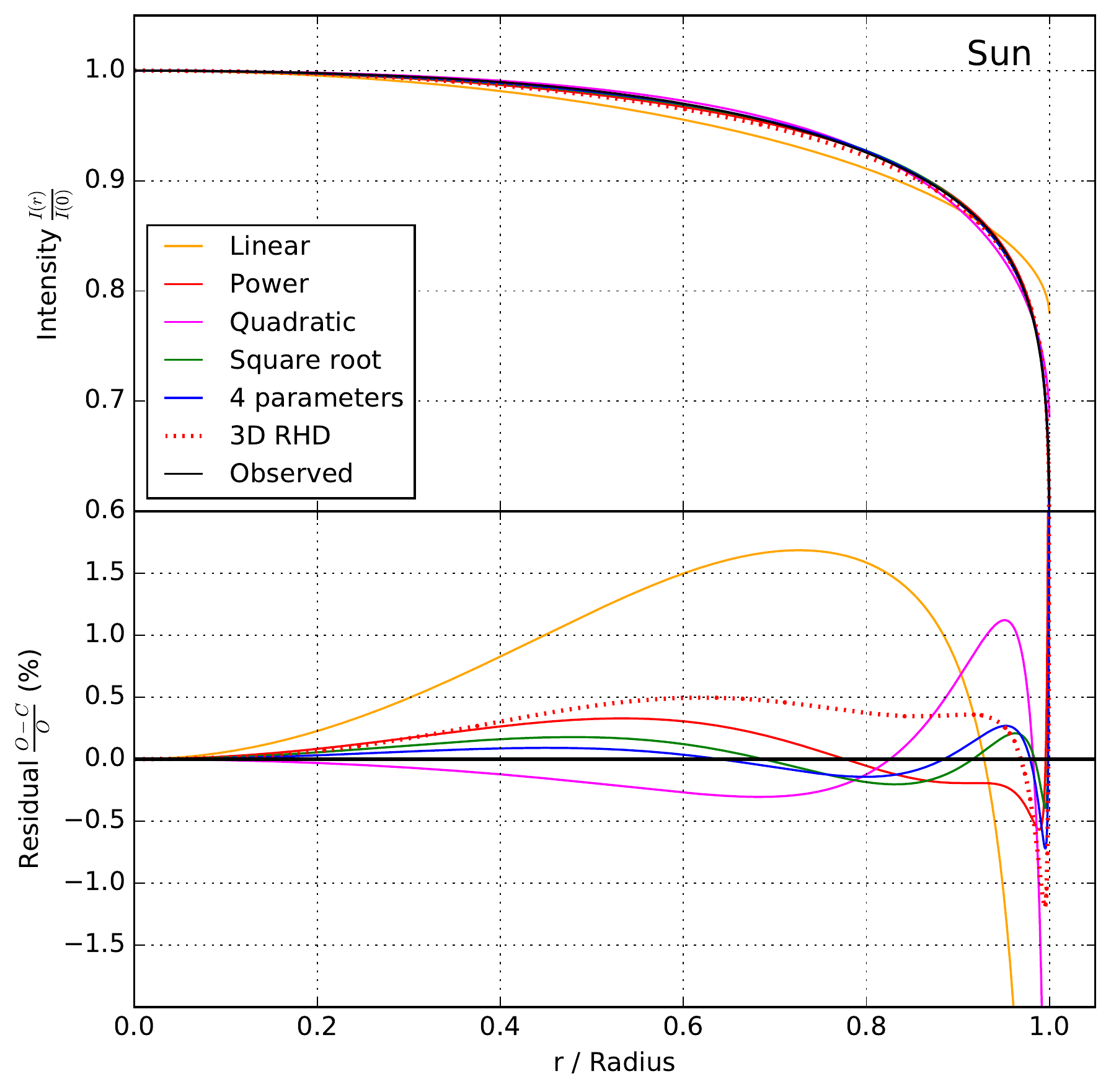}
        \caption{Comparison of different parametric limb darkening models of the Sun with the observed limb darkening profile measured by \citetads{1977SoPh...52..179P} in the $H$ band.
        The residuals in percentage of the observed intensity profile are shown in the lower panel.
        \label{solarprofiles}}
\end{figure}
The scaled solar model is also useful owing to the generally poor realism of the single and two-parameter LD approximations.
We show in Fig.~\ref{solarprofiles} a comparison of different approximate parametric models adjusted to the observed $H$-band solar intensity profile (Eq.~\ref{solarmodel}) by \citetads{1977SoPh...52..179P}.
We also show the prediction of the solar intensity profile from our 3D convective model (see Sect.~\ref{3Dmodel}) for comparison.
It is clear from this diagram that the linear and quadratic parametric models are poor representations of the actual intensity profile of the Sun.
This was also the conclusion reached, for example, by \citetads{2007ApJ...656..483H}, \citetads{2015A&A...584A..80M} and \citetads{2016MNRAS.457.3573E}.
The square root model and the four-parameter law introduced by \citetads{2000A&A...363.1081C} appear to be better approximations.
A specific difficulty with the square root model is the high degree of correlation between its two parameters $c$ and $d$ due to the relative similarity of the linear (for parameter $a$) and square root (for parameter $b$) functions that it combines.
But both the square root and four-parameter laws are not significantly more accurate approximations of the profile of the Sun than the single-parameter power law introduced by \citetads{1997A&A...327..199H}, at least in the infrared $H$ band that we discuss here.

\begin{figure*}
        \centering
        \includegraphics[width=17cm]{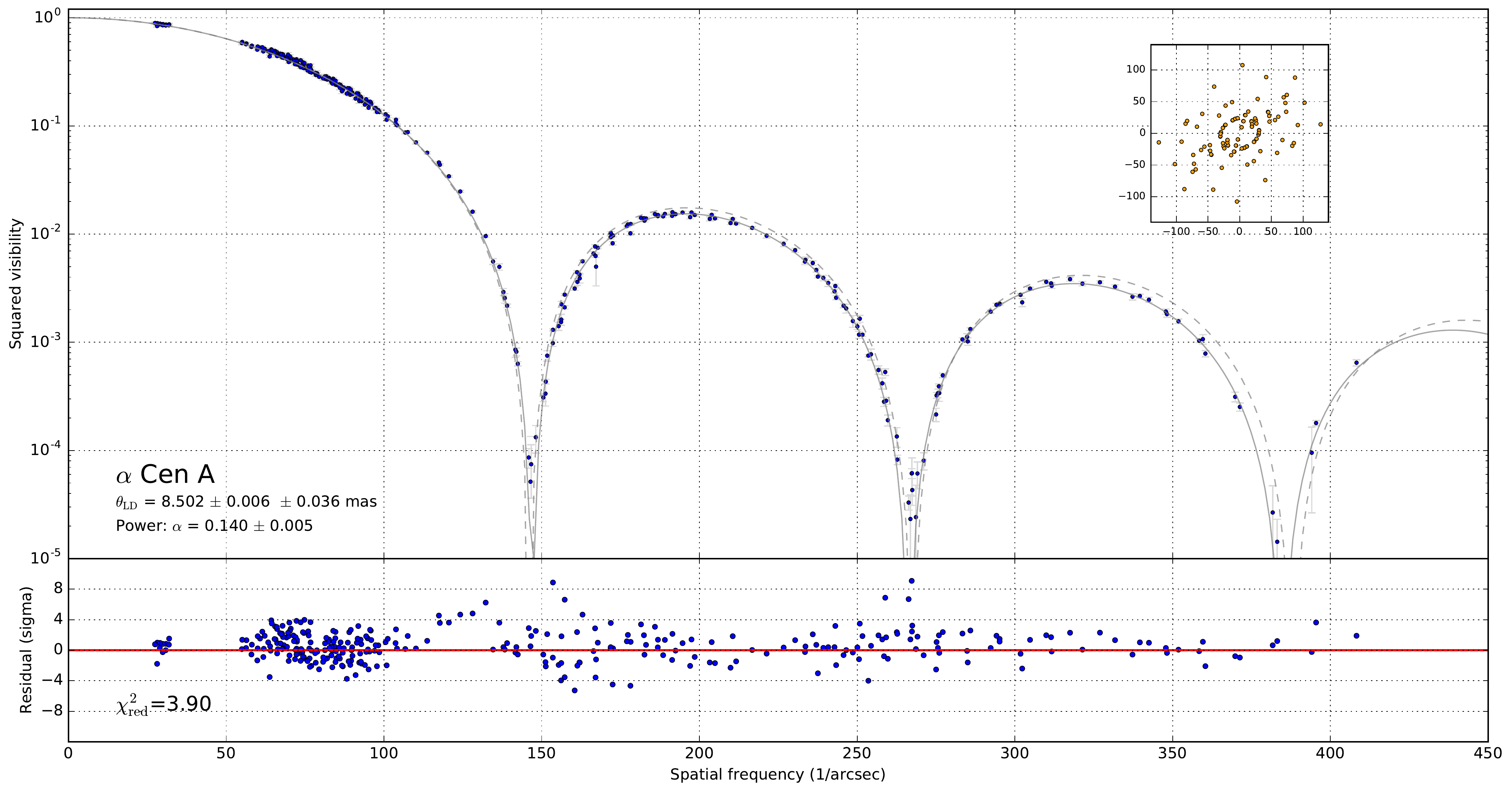}
        \caption{Adjustment of a power law limb darkened disk model to the PIONIER squared visibilities of $\alpha$\,Cen~A (solid gray curve). The dashed gray curve represents the best-fit uniform disk model.
        The bottom panels show the residuals of the fit in number of times the statistical error bar.
        The coverage of the $(u,v)$ plane is shown in the upper right corner.
        \label{ldfitA}}
\end{figure*}

\begin{figure*}
        \centering
        \includegraphics[width=17cm]{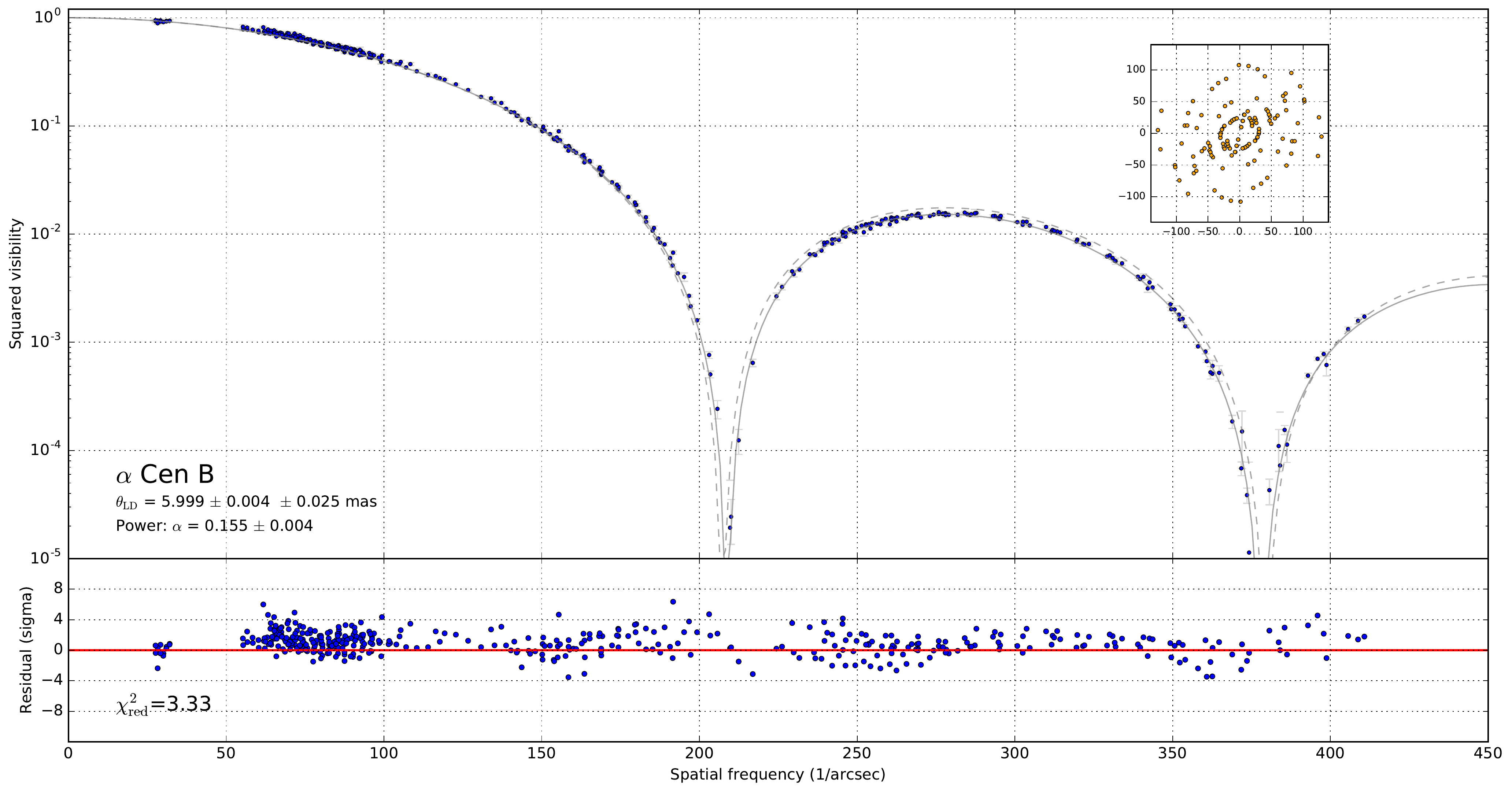}
        \caption{Power law limb darkened disk model fit and residuals for $\alpha$\,Cen~B (same caption as Fig.~\ref{ldfitA}).
        \label{ldfitB}}
\end{figure*}

The results of the fits of different parametric LD models, together with the 3D atmosphere model presented in Sect.~\ref{3Dmodel}, are listed in Table~\ref{LDmodels}, and the corresponding residuals are shown in Fig.~\ref{residualsA} and \ref{residualsB}.
A discussion of the residuals of the different models is presented in Sect.~\ref{LDquality}.
The best-fit power law models adjusted to the PIONIER squared visibilities of $\alpha$\,Cen A and B are presented in Fig.~\ref{ldfitA} and \ref{ldfitB}, respectively.

\begin{sidewaystable*}
        \caption{Uniform and limb darkened disk model parameters for $\alpha$\,Cen A, $\alpha$\,Cen B and the Sun.}
        \centering          
        \label{LDmodels}
        \begin{tabular}{llcclccl}
	\hline\hline
        \noalign{\smallskip}
	       & \multicolumn{3}{c}{$\alpha$\,Cen A (G2V)} & \multicolumn{3}{c}{$\alpha$\,Cen B (K1V)} & Sun (G2V) \\
\hline 
\noalign{\smallskip}
        LD model & LD parameters & $\theta_\mathrm{LD}(A)$ [mas]$^a$ & $\chi^2_\mathrm{red}$ & LD parameters & $\theta_\mathrm{LD}(B)$ [mas]$^a$ & $\chi^2_\mathrm{red}$ & Observed LD \\
        \noalign{\smallskip}
         \hline
        \noalign{\smallskip}
Uniform &  $-$ & $8.347 \pm 0.004 \pm 0.035$ & 15.23 & $-$ & $5.883 \pm 0.003 \pm 0.025$ & 14.26 & $-$ \\
\hline \noalign{\smallskip}
Linear (fixed)$^b$  & $u = 0.2392$ (fixed) & $8.505 \pm 0.003 \pm 0.036$ & 5.50 & $u = 0.2698$ (fixed) & $6.003 \pm 0.002 \pm 0.025$ & 4.69 & $u_\odot=0.2227 \pm 0.0005$ \\
\hline \noalign{\smallskip}
Linear (fit) & $u = 0.1761 \pm 0.0062$ & $8.458 \pm 0.005 \pm 0.035$ & 4.24 & $u = 0.1907 \pm 0.0048$ & $5.962 \pm 0.003 \pm 0.025$ & 2.89 & $u_\odot=0.2227 \pm 0.0005$ \\
\hline \noalign{\smallskip}
Quadratic & $a = +0.191 \pm 0.026$ & $8.451 \pm 0.013 \pm 0.035$ & 4.25 &  &  &  & $a_\odot=+0.0908 \pm 0.0004$ \\
 & $b = -0.031 \pm 0.054$ &  &  &  & & & $b_\odot=+0.2309 \pm 0.0007$ \\
\hline \noalign{\smallskip}
Square root & $c = +0.29 \pm 0.15$ & $8.446 \pm 0.017 \pm 0.035$ & 4.24 &  &  &  & $c_\odot=-0.2257 \pm 0.0004$  \\
 & $d = -0.19 \pm 0.25$ &  &  &  &  & & $d_\odot=+0.7230 \pm 0.0006$  \\
\hline \noalign{\smallskip}
4 param. (fixed)$^c$  & $a_1 = +0.7127$ & $8.540 \pm 0.003 \pm 0.036$ & 4.98 & $a_1 = +0.7353$ & $6.030 \pm 0.002 \pm 0.025$ & 3.93 & $a_{1\,\odot}=+0.9789 \pm 0.0069$ \\
 & $a_2 = -0.0452$ &  &  & $a_2 = -0.0144$ & & & $a_{2\,\odot}= -0.9644 \pm 0.0169$ \\
 & $a_3 = -0.2643$ &  &  & $a_3 = -0.2485$ & & & $a_{3\,\odot}= +0.8481 \pm 0.0174$  \\
 & $a_4 = +0.1311$ &  &  & $a_4 = +0.1083$ & & & $a_{4\,\odot}= -0.3371 \pm 0.0065$  \\
\hline \noalign{\smallskip}
3D & $-$ & $8.534 \pm 0.003 \pm 0.036$ & 4.85 & $-$ & $6.037 \pm 0.002 \pm 0.025$ & 4.40 & $-$ \\
\hline \noalign{\smallskip}
Power & $\alpha = 0.1404 \pm 0.0050$ & $8.502 \pm 0.006 \pm 0.036$ & 3.90 & $\alpha = 0.1545 \pm 0.0044$ & $5.999 \pm 0.004 \pm 0.025$ & 3.33 & $\alpha_\odot=0.15027 \pm 0.00006$ \\
\hline \noalign{\smallskip}
Scaled solar & $s_0 = 0.716 \pm 0.042$ & $8.498 \pm 0.007 \pm 0.036$ & 4.38 & $s_0 = 0.598 \pm 0.027$ & $5.996 \pm 0.004 \pm 0.025$ & 3.13 & $s_{0\,\odot} = 0.59045$ $^d$ \\
    	\hline
        \end{tabular}
        \tablefoot{
        $^a$ The statistical and systematic uncertainties on the angular diameters are listed separately as $\pm \sigma_\mathrm{stat} \pm \sigma_\mathrm{syst}$.
        $^b$ The linear limb darkening coefficients $u$ are taken from \citetads{2011A&A...529A..75C}.
	$^c$ The four-parameter non linear limb darkening coefficients $a_1...a_4$ are taken from \citetads{2011A&A...529A..75C}.
        $^d$ The value of $s_0$ for the Sun is the average over the $H$ band from the observed profile by \citetads{1977SoPh...52..179P}.
        }
\end{sidewaystable*}

\subsubsection{Three-dimensional atmosphere model\label{3Dmodel}}

We interpret our PIONIER observations with the result from realistic 3D radiative hydrodynamical simulations of convection.
We used the {\scriptsize STAGGER} code (Nordlund \& Galsgaard 1995\footnote{\url{http://www.astro.ku.dk/~kg/Papers/MHD_code.ps.gz}}; \citeads{2012A&A...539A.121B}) which was previously used to interpret interferometric angular diameter determinations (e.g. \citeads{2006A&A...446..635B}; \citeads{2011A&A...534L...3B}).
These state-of-the-art simulations provide extremely realistic modeling of the solar surface (see e.g.~\citeads{1998ApJ...499..914S}, \citeads{2009LRSP....6....2N}) from first principles without the need of tuned parameters (e.g., mixing-length).
These simulations also provide in principle reliable limb darkened intensities.  
The code solves the full set of conservative hydrodynamical equations coupled to an accurate treatment of the radiative transfer.
The equations are solved on a staggered mesh with a sixth order explicit finite difference scheme.
We used the 3D models for $\alpha$\,Cen~A and B obtained by \citetads{2008MmSAI..79..670B} and \citetads{2006A&A...446..635B}.
The domains of simulations are local boxes at the surface  ($6 \times 6 \times 5.9$\,Mm for A, and $6 \times 6 \times 5.7$\,Mm for B).
They contain the entropy minima  and are extended deep enough to have a flat entropy profile at the bottom (adiabatic regime).
The code uses periodic boundary conditions horizontally and open boundaries vertically. At the bottom of the simulation, the inflows have constant entropy and pressure.
A realistic equation-of-state  accounts for ionization, recombination, and dissociation \citepads{1990ApJ...350..300M} and continuous line opacities \citepads{2008A&A...486..951G}.
Radiative transfer is solved using the Feautrier scheme along several vertical and inclined rays.
The wavelength dependence of the radiative transfer is taken into account using a binning scheme in which the monochromatic lines are collected into 12 bins. 
The stellar parameters that define our 3D models are: mean $T_\mathrm{eff} = 5820$\,K, $\log g = 4.32$, $[\mathrm{Fe/H}] = +0.25$ dex for $\alpha$\,Cen~A, and mean $T_\mathrm{eff} = 5240$\,K, $\log g = 4.51$, $[\mathrm{Fe/H}] =+0.25$ dex for $\alpha$\,Cen~B. 

\citetads{2006A&A...446..635B} presented a first comparison of only two VLTI/VINCI visibility measurements obtained in the second lobe of the visibility function of $\alpha$\,Cen~B with the prediction of 3D hydrodynamical simulation.
Since the difference between 3D and 1D  is modest for a K dwarf, especially in the $K$ band, we could only conclude that the 3D approach gave a marginally better fit than the classical 1D approach. 
Now we have a much better coverage of the visibility function with PIONIER. Our 3D determinations of the angular diameters are (Table~\ref{LDmodels}) $\theta_{\rm 3D}[A] = 8.534 \pm 0.003$\,mas ($\chi^2_{\rm red}=4.85$) and $\theta_{\rm 3D}[B] = 6.037 \pm 0.002$ mas ($\chi^2_{\rm red}=4.40$).
In these error bars we ignored the contribution of the uncertainty in wavelength that is a simple scaling factor common to both stars.
Calculating equivalent 1D, LD angular diameters using non-linear four-parameter LD approximations tabulated by \citetads{2011A&A...529A..75C}, we obtain $\theta_{\rm 1D}[A] = 8.540 \pm 0.003$\,mas  ($\chi^2_{\rm red}=4.98$) and $\theta_{\rm 1D}[B] = 6.030 \pm 0.002$\,mas ($\chi^2_{\rm red}=3.93$), which is comparable to the 3D LD values.
We note that our PIONIER diameter of $\alpha$ Cen B ($\theta_{\rm 3D}[B] = 6.037 \pm 0.002 \pm 0.025 $ mas) is within $1\sigma$ of the value derived from VINCI observations by \citeads{2006A&A...446..635B} ($\theta_{\rm 3D\ VINCI}[B] = 6.000 \pm 0.021 $ mas). 

\subsubsection{Quality of the limb darkening models\label{LDquality}}

\begin{figure}[]
        \centering
        \includegraphics[width=\hsize]{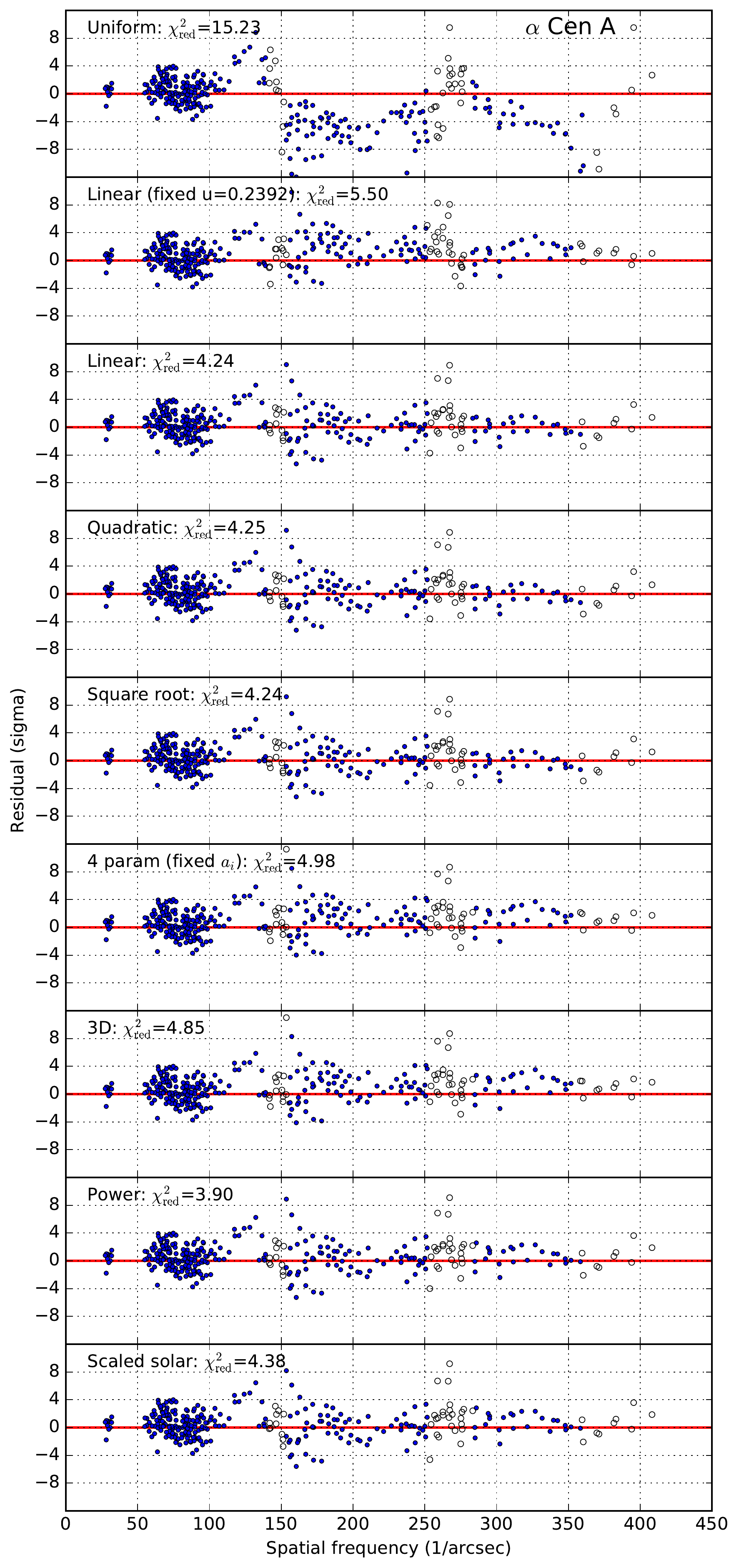}
        \caption{Residuals (observed - predicted) of the fits of different limb darkening models to the squared visibilities of $\alpha$\,Cen A.
        The open circles indicate $V^2$ residuals close to the minima of the visibility function ($V^2 < 10^{-3}$) for which the fit is potentially more unstable.
        The fixed linear and 4-parameter LD models assume coefficient values from \citetads{2011A&A...529A..75C}.
        \label{residualsA}}
\end{figure}

\begin{figure}[]
        \centering
        \includegraphics[width=\hsize]{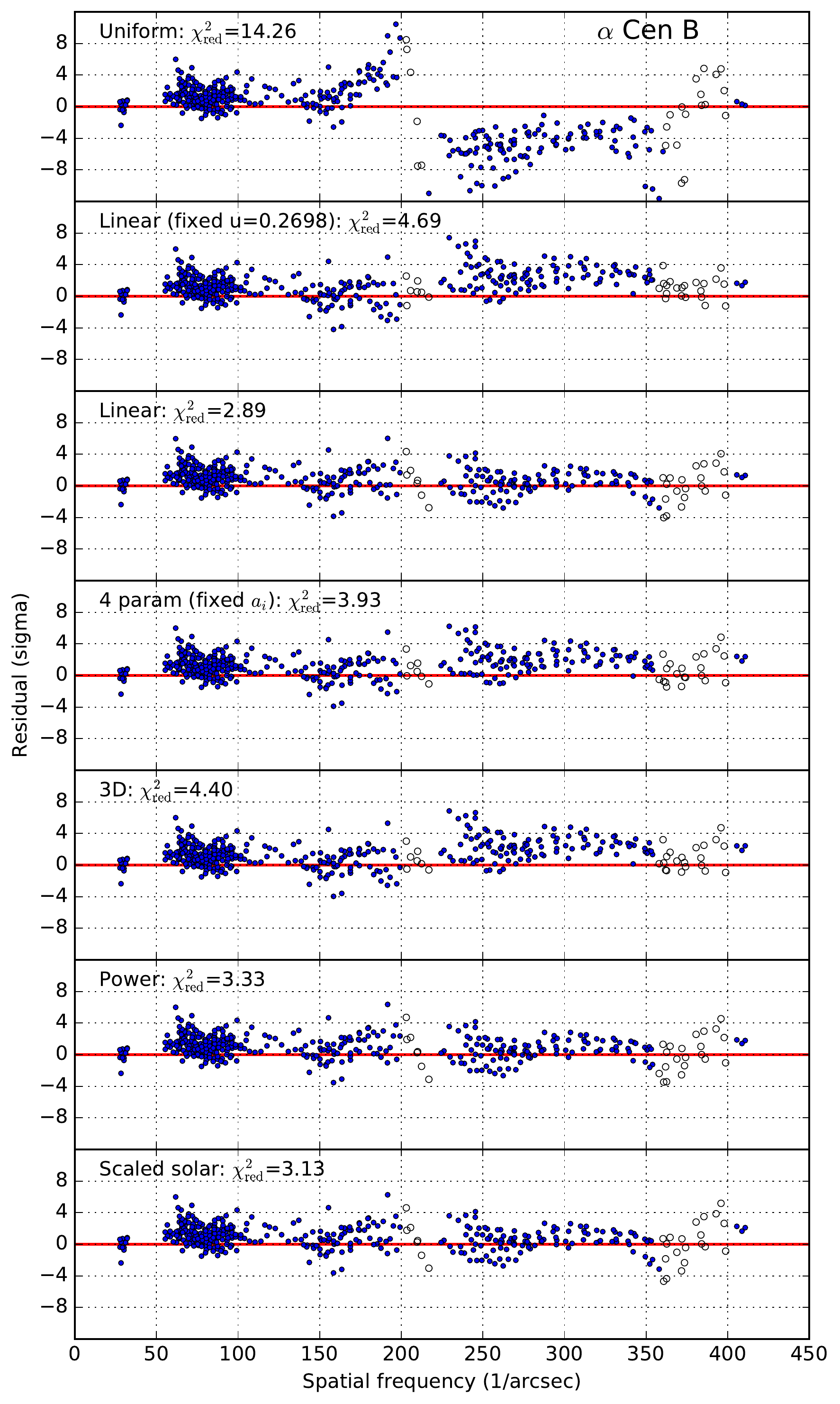}
        \caption{Residuals (observed - predicted) of the fits of different limb darkening models to the squared visibilities of $\alpha$\,Cen B  (same caption as Fig.~\ref{residualsA}).
        \label{residualsB}}
\end{figure}

An indication of the quality of the LD model fit on the PIONIER visibilities is given by the minimum reduced $\chi^2$ value (Figs.~\ref{residualsA} and \ref{residualsB}), but this is not a perfect indicator as we typically have many more data points in the first lobe of the visibility function than in the upper lobes.
This results in a high weight in the $\chi^2$, which does not fully reflect the quality of the LD parameter fit, as the higher order lobes of the visibility function are the lobes that constrain these parameters ($u, \alpha, a, b,...$).
We therefore discuss here the properties of the residuals shown in Fig.~\ref{residualsA} and Fig.~\ref{residualsB}.

As expected, the uniform disk model is excluded as it largely overestimates the contrast of the second and higher order lobes of the visibility functions.

The single-parameter linear, power law and scaled solar LD models show very similar residuals.
The quality of the fit is very good for the three types of models for $\alpha$\,Cen A, with essentially symmetric residuals around zero for the second and third lobes of the visibility function.
For $\alpha$\,Cen B the linear limb darkening model (fitting $u$ as a variable) results in a slightly lower $\chi^2$ value than the power law and scaled solar models.
But we sample only the first and second lobes of the visibility function for this star, and we are therefore insensitive to higher order deviations between the LD model and the observed profile.
In other words, our limited angular resolution of the stellar disk of $\alpha$\,Cen~B does not allow us to discriminate between the detailed shape of the intensity profile of these three models.
The LD angular diameters of both $\alpha$\,Cen~A and B are very close for the power law and scaled solar LD models, with a maximum difference between them of less than 0.1\%.
This agreement is expected as the LD angular diameter is essentially constrained by the position in spatial frequency of the minima of the visibility function, which are only mildly affected by the exact shape of intensity profile.

The two-parameter quadratic and square root models provide a very good fit to the observed visibility distributions for $\alpha$\,Cen~A.
These models cannot be adjusted to star B because the angular resolution is too limited.
The fit residuals are indistinguishable from each other and from the single parameter models (linear and power law).
We conclude that the additional parameter of the quadratic and square root models does not provide a significant advantage compared to single-parameter models, at the level of angular resolution we achieved on $\alpha$\,Cen~A.

The four-parameter models with fixed coefficients taken from \citetads{2011A&A...529A..75C} overestimate the LD of both stars A and B and therefore also overestimate their angular diameters.
We cannot fit the four model parameters $a_i$ simultaneously as this would require that we resolve the stars up to at least the fifth lobe of their visibility function. We are therefore limited to a comparison of the model predictions with the data.

The 3D hydrodynamical model has no LD parameter to adjust as its properties are set by the underlying physics described in Sect.~\ref{3Dmodel}.
We observe an overestimation of the LD of both $\alpha$\,Cen~A and B by their respective 3D LD models, with residuals very similar to the four-parameter model.

In conclusion, from the comparison of the selected parametric LD models to the visibility measurements of $\alpha$\,Cen~A and B, we find that the single parameter LD models (linear, power law and scaled solar) provide a satisfactory representation of the observations.
Models with one additional parameter (quadratic, square root) or fixed LD models (four-parameter and 3D) do not result in a significant improvement of the quality of the fits.
We showed in Sect.~\ref{parametric} (Fig.~\ref{solarprofiles}) that the observed solar LD profile is poorly reproduced  by the linear and quadratic LD models, while the square root, four-parameter, power law, scaled solar and 3D models provide a better match to the data.

Taking into account both our observations of $\alpha$\,Cen A and B and the LD profile of the Sun, we therefore conclude that the power law and scaled solar intensity profiles represent the optimum compromise between the number of model parameters and the fidelity to the actual LD profile.
For the following discussion, we therefore adopt as estimates of the photospheric angular diameters the best-fit angular diameters from the power law LD model (Figs.~\ref{ldfitA} and \ref{ldfitB}).
This choice has the advantage to allow a straightforward, single-parameter comparison with the LD of the Sun and other types of stars (see also the discussion by \citeads{1997A&A...327..199H}).

\subsection{Search for spots and companions \label{companions}}

\begin{figure}[]
        \centering
        \includegraphics[width=8cm]{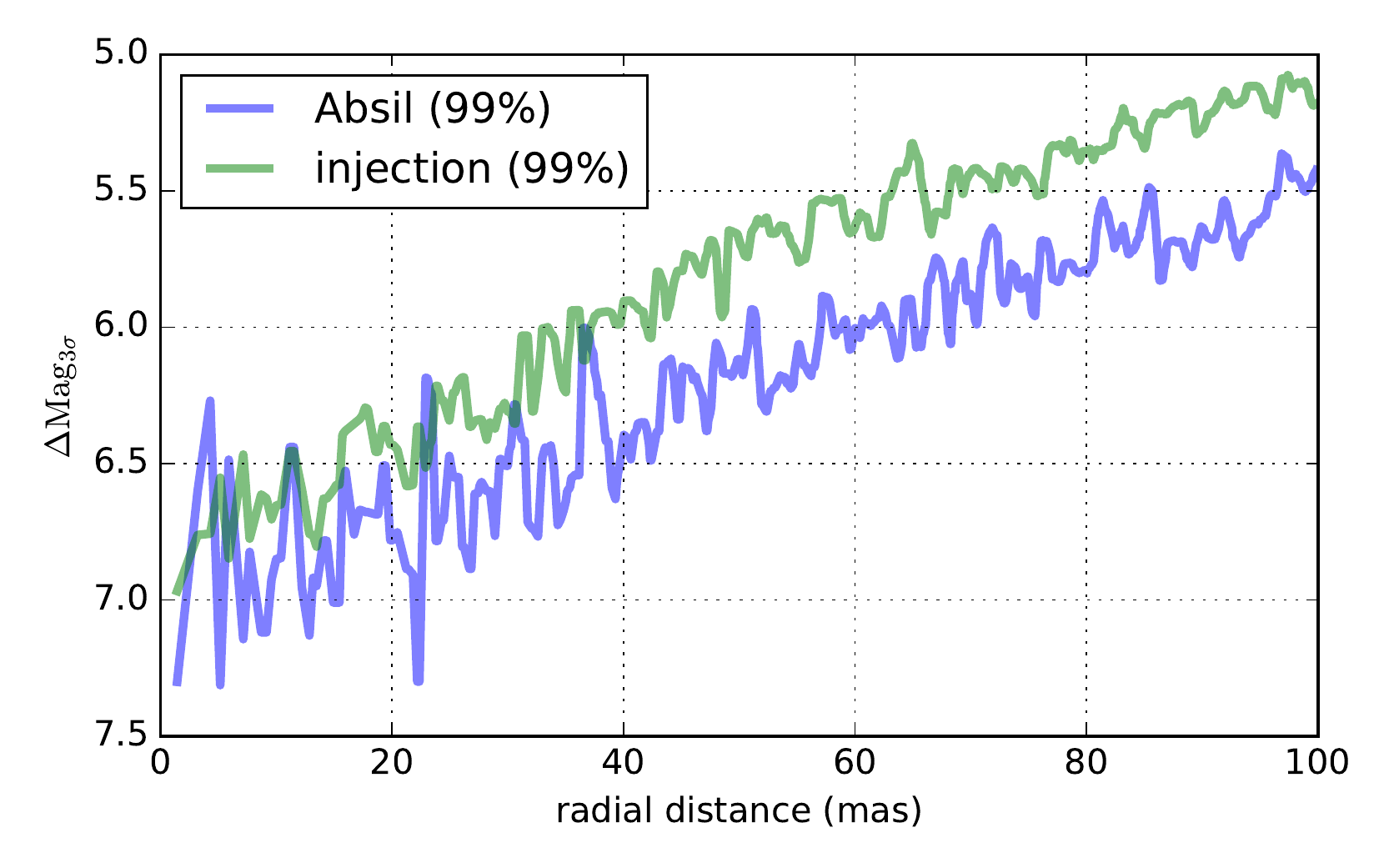}
        \includegraphics[width=8cm]{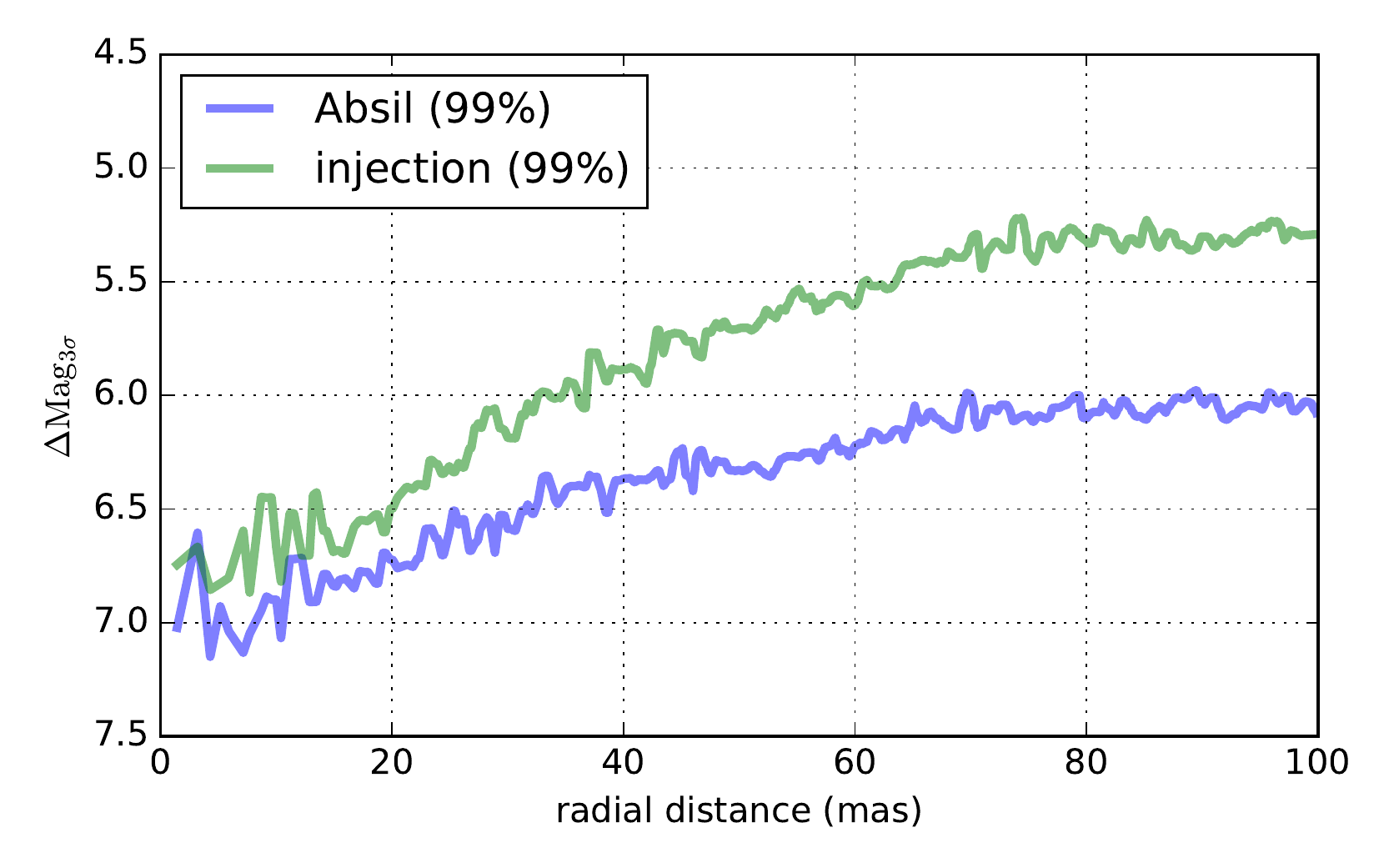}
        \caption{Contrast limits in the $H$ band at $3\sigma$ for the presence of companions within 100\,mas of $\alpha$\,Cen A (top) and B (bottom).
        The blue curves correspond to the sensitivity limits as defined by \citetads{2011A&A...535A..68A} and the green curves for the prescription by \citetads{2015A&A...579A..68G}.
        \label{contrasts}}
\end{figure}

Owing to the efficiency of the Earth rotation supersynthesis for $\alpha$\,Cen A and B, the coverage of the $(u,v)$ plane of spatial frequencies (sub-panels in Fig.~\ref{ldfitA} and \ref{ldfitB}) is sufficiently good to allow us to search for the presence of additional sources in their close environment (companions); we can also search for asymmetries on their photospheres due to stellar spots, which can be brighter or darker than the average photosphere.
The two stars have coronal cycles in the X-ray and ultraviolet domains with periods around 19 and 8 years, respectively, as shown by \citetads{2014AJ....147...59A, 2015AJ....149...58A}.
The presence of spots on the surface of the stars could affect the measured interferometric fringe visibilities and therefore bias the estimates of the angular diameter and LD parameters.
We employed the {\tt CANDID} tool \citepads{2015A&A...579A..68G} to search simultaneously the closure phases, squared visibilities and closure amplitudes of PIONIER for the signature of additional point sources within 100\,mas from the centers of the two stars.
To prevent the displacement of companions or spots, we only considered the data sets obtained on 29 and 30 May 2016.
No significant secondary source over $3\sigma$ is detected for $\alpha$\,Cen~A over the interferometric field of view.
The brightest candidate companion (outside of the photosphere) is found at a flux ratio $f=0.14\%$ (with respect to $\alpha$\,Cen~A) and is not statistically significant ($2.2\sigma$).
Within the apparent disk of the star, there is no candidate spot above the noise level ($f=0.07\%$).
For $\alpha$\,Cen~B, the brightest candidate companion has $f=0.16\%$ but is not statistically significant ($2.5\sigma$).
A candidate spot is found at $f=0.21\%$ but is not statistically significant either ($1.8\sigma$).
A plot of the sensitivity limit as a function of separation from the center of the stellar disk and expressed in terms of contrast in the $H$ band is presented in Fig.~\ref{contrasts}.

From this analysis, we exclude the presence of isolated spots on the surfaces of both stars at a contrast level of at least $\Delta H = 6.5$\,mag, corresponding to a relative flux contribution of $0.25\%$ with respect to the total stellar flux.
This implies that the limb darkening coefficients determined in Sect.~\ref{LDmodelfit} are not significantly biased by spots.

\section{Discussion\label{discussion}}

\subsection{Comparison with literature models and the Sun \label{comparisonLD}}

The simplest comparison between the literature and our measurements is for the single-parameter linear LD coefficient $u$.
Although it is not a good representation of the true intensity profile of the star (Sect.~\ref{LDquality}), its broad usage in the literature makes it a useful basis of comparison with our measurements.
A sample of values of $u$ from the literature is presented in Table~\ref{LDliterature}.
To read the tables from the different authors, we adopt the following turbulence and metallicity parameters for $\alpha$\,Cen A and B from \citetads{2015A&A...582A..81J} and \citetads{2015A&A...582A..49H}:
$T_\mathrm{eff}[A] = 5792 \pm 16$\,K, $\xi_\mathrm{turb}[A] = 1.20 \pm 0.07$\,km\ s$ ^{-1}$, $[\mathrm{Fe/H}][A]= +0.24$, and 
$T_\mathrm{eff}[B] = 5231 \pm 20$\,K, $\xi_\mathrm{turb}[B] = 0.99 \pm 0.31$\,km\ s$ ^{-1}$, $[\mathrm{Fe/H}][B] = +0.22$.
These parameters are generally in good agreement with the following spectroscopic determination by \citetads{2008A&A...488..653P}:
$T_\mathrm{eff}[A] = 5847 \pm 27$\,K, $\xi_\mathrm{turb}[A] = 1.46 \pm 0.03$\,km\ s$ ^{-1}$, $[\mathrm{Fe/H}][A]= +0.24 \pm 0.03$, and 
$T_\mathrm{eff}[B] = 5316 \pm 28$\,K, $\xi_\mathrm{turb}[B] = 1.28 \pm 0.15$\,km\ s$ ^{-1}$, $[\mathrm{Fe/H}][B] = +0.25 \pm 0.04$.
We adopt the masses determined by \citetads{2016kervella} as follows: $m_A  = \massA \pm \massAerr\,M_\odot$ and $m_B = \massB \pm \massBerr\,M_\odot$.
The surface gravity parameter of $\alpha$\,Cen~A and B can be deduced from the combination of these masses and our radius measurements, and we obtain $\log g[A] = \loggA \pm \loggAerr$ and $\log g[B] = \loggB \pm \loggBerr$, considering the IAU solar mass conversion constant of $\GMnom = 1.3271244\ 10^{20}$\,m$^3$\,s$^{-2}$ \citepads{2016AJ....152...41P}.
Our new values are within $1\sigma$ of the spectroscopic estimates from \citetads{2008A&A...488..653P} ($\log g[A] = 4.34 \pm 0.12$, $\log g[B] = 4.44 \pm 0.15$) and in perfect agreement with the calibration of \emph{Gaia} benchmark stars by \citetads{2015A&A...582A..49H} ($\log g[A] = 4.31 \pm 0.01$, $\log g[B] = 4.53 \pm 0.03$).
They are however one to two orders of magnitude more accurate thanks mainly to the improved masses from \citetads{2016kervella}.

We did not interpolate the parametric LD tables from the literature, that is, we directly chose the closest model parameters to the values listed above for the $H$ band.
The reason for this choice is that the dependence of the linear limb darkening coefficient $u$ is potentially a complex and non-linear function of the model parameters ($T_\mathrm{eff}, \log g,...$), and correlations may exist between them.
A simple interpolation considering each parameter separately may therefore introduce artefacts, and a more complex multi-dimensional interpolation taking the correlations into account is beyond the scope of the present comparison. 
This is however not a real limitation in practice, as the atmosphere model grids are densely populated around the solar parameters.
The parameters employed to read the tabulated values of $u$ are therefore close to the true values for the different models.

All model values in the literature systematically overestimate $u$ for both $\alpha$\,Cen A and B.
Many of these coefficients are based on the same atmosphere models (usually ATLAS9; \citeads{1979ApJS...40....1K}; \citeads{2004astro.ph..5087C}), and therefore share the same underlying numerical basis.
But the STAGGER 3D simulations \citepads{2015A&A...573A..90M} also show the same trend.
We also observe a significantly larger discrepancy between our measured value of $u$ and the spherical atmosphere model by \citetads{2013A&A...556A..86N} than with the planar version from the same authors.
A plausible explanation for this difference is that a linear model is a too crude representation of the limb darkening profile predicted by a spherical model, as the intensity drops sharply near the limb ($\mu \approx 0$).
Depending on the adopted weighting in the linear model fit to the exact spherical profile, this usually results in an overestimation of $u$.
Overall, the level of discrepancy between our measurements and the published models is such that none of them provides a prediction for $u$ in agreement with the observed values within their error bars.
This is a surprising result considering that $\alpha$\,Cen A and B, and component A in particular, are similar to the Sun, for which one would expect very well-calibrated atmosphere models.

We note that the method used to derive parametric LD models from stellar atmosphere models may not be well suited for interferometric observations.
The wavelength dependent intensity profiles produced by atmosphere models are usually computed with constant steps in $\mu$.
The resulting curve $I(\mu)/I(1)$ is then approximated by a parametric model (usually polynomial) by giving a uniform weight to all points of the $I(\mu)$ function.
But interferometry sees the star not as a 1D radial cut, but as a circular disk.
More precisely, it measures the visibility as the Hankel transform of the intensity profile.
The effective weighting of the profile in the visibility function is thus different from the approximate polynomial fit of the original atmosphere model.
As a result, the visibility function of a polynomial LD model often differ substantially from the visibility function computed using the original atmosphere model directly.
Using the non-approximated model outputs (e.g.~from ATLAS9) instead of parametric models would be a step in the right direction to obtain a better match to the observed intensity profiles.
But the fact that the non-approximated 3D models significantly overestimate the LD of $\alpha$\,Cen and the Sun (as the parametric models) indicates that the main problem lies in the underlying atmosphere physics, and not simply in the polynomial approximation of the numerical model.

\citetads{1998A&A...333..338H} derived empirical power law coefficients $\alpha$ between 0.13 and 0.17 from the solar LD observations of reference by \citetads{1977SoPh...52..179P} over the $H$ band, and we confirm an integrated value of $\alpha_\odot = 0.15027 \pm 0.00006$ (Table~\ref{LDmodels}).
The solar power law exponent is therefore very comparable to the values we obtain for $\alpha$\,Cen A and B: $\alpha(A) = 0.1404 \pm 0.0050$ and $\alpha(B) = 0.1545 \pm 0.0055$.
Hence, our interferometric LD measurements show that the LD of $\alpha$\,Cen~A is slightly lower than that of the Sun, while the LD of $\alpha$\,Cen~B is statistically identical to the solar value (Fig.~\ref{profiles}).
For comparison, the power law exponent $\alpha = 0.23 \pm 0.05$ found by \citetads{2010A&A...517A..64M} in the $K$ band for the red subgiant $\eta$\,Ser (K0III-IV) is higher than the power law exponent we find for $\alpha$\,Cen~B in the $H$ band. $\eta$\,Ser is cooler  than $\alpha$\,Cen~B and its effective gravity is significantly lower ($T_\mathrm{eff}=4955$\,K, $\log g = 3.2$; \citeads{2007A&A...475.1003H}).
\citetads{2008A&A...485..561L} observed the even cooler and larger K0III giant Arcturus ($T_\mathrm{eff} \approx 4300$\,K, $\log g = 1.5$) in the $H$ band and measured $\alpha = 0.258 \pm 0.003$ to be also significantly higher than the value of $\alpha$\,Cen~B.

\begin{figure}[]
        \centering
        \includegraphics[width=\hsize]{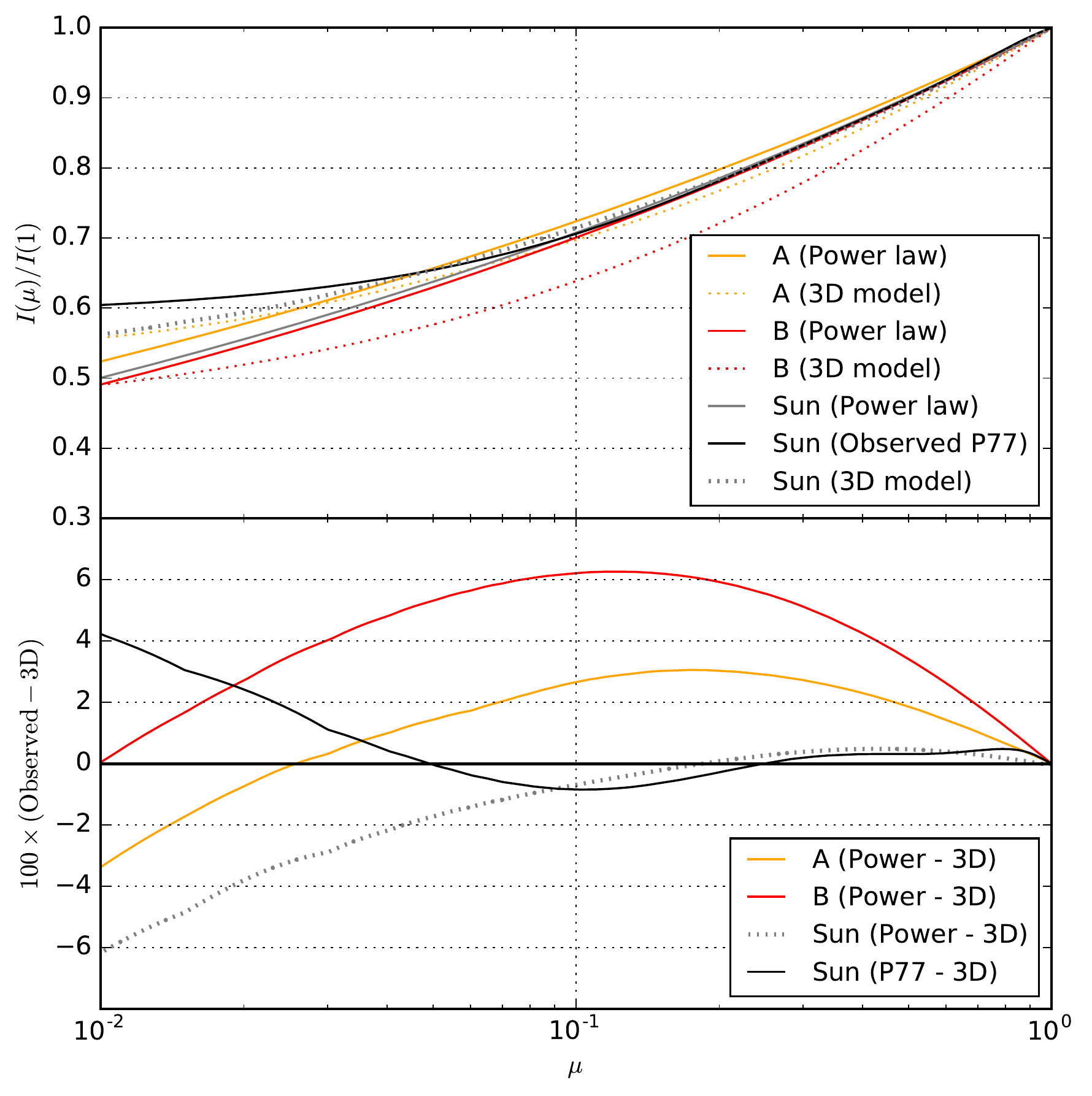}
        \caption{Intensity profiles $I(\mu)/I(1)$ for $\alpha$\,Cen~A, $\alpha$\,Cen~B and the Sun.
        {\it Top panel:} The observed power laws $I(\mu)/I(1) = u^\alpha$ are represented with solid curves, together with the measured polynomial by \citetads{1977SoPh...52..179P} (P77, black). The 3D model predictions are represented with dashed curves.
        {\it Bottom panel:} Differences between the observed profiles and the 3D model predictions.
        \label{profiles-comparison}}
\end{figure}

A disagreement between observations and models also exists for the Sun.
\citetads{2013A&A...554A.118P} compared the predictions of model atmospheres of the Sun to the \citetads{1977SoPh...52..179P} observations in the near-infrared.
In line with our observations, they find a consistent overestimation of the LD for 1D MARCS and PHOENIX models in the infrared, and particularly in the $H$ band, while the comparison with 3D models shows a better (although not perfect) agreement (their Fig.~3).
A comparison of the 3D model predictions with the observed intensity profiles of $\alpha$\,Cen and the Sun is presented in Fig.~\ref{profiles-comparison}.
The 3D model predictions are in good agreement with the observed profile for $\mu > 0.1$, but differ significantly for lower values.
The overestimation of the solar limb darkening by our 3D model prediction is also visible in Fig.~\ref{solarprofiles}.

A possible explanation for this difference is the inaccurate treatment of the opacity due to the hydrogen anion (see e.g.~\citeads{2014LRSP...11....2P}).
The fact that $\alpha$\,Cen~B is magnetically active may also play a role.
%

\begin{figure}[]
        \centering
        \includegraphics[width=\hsize]{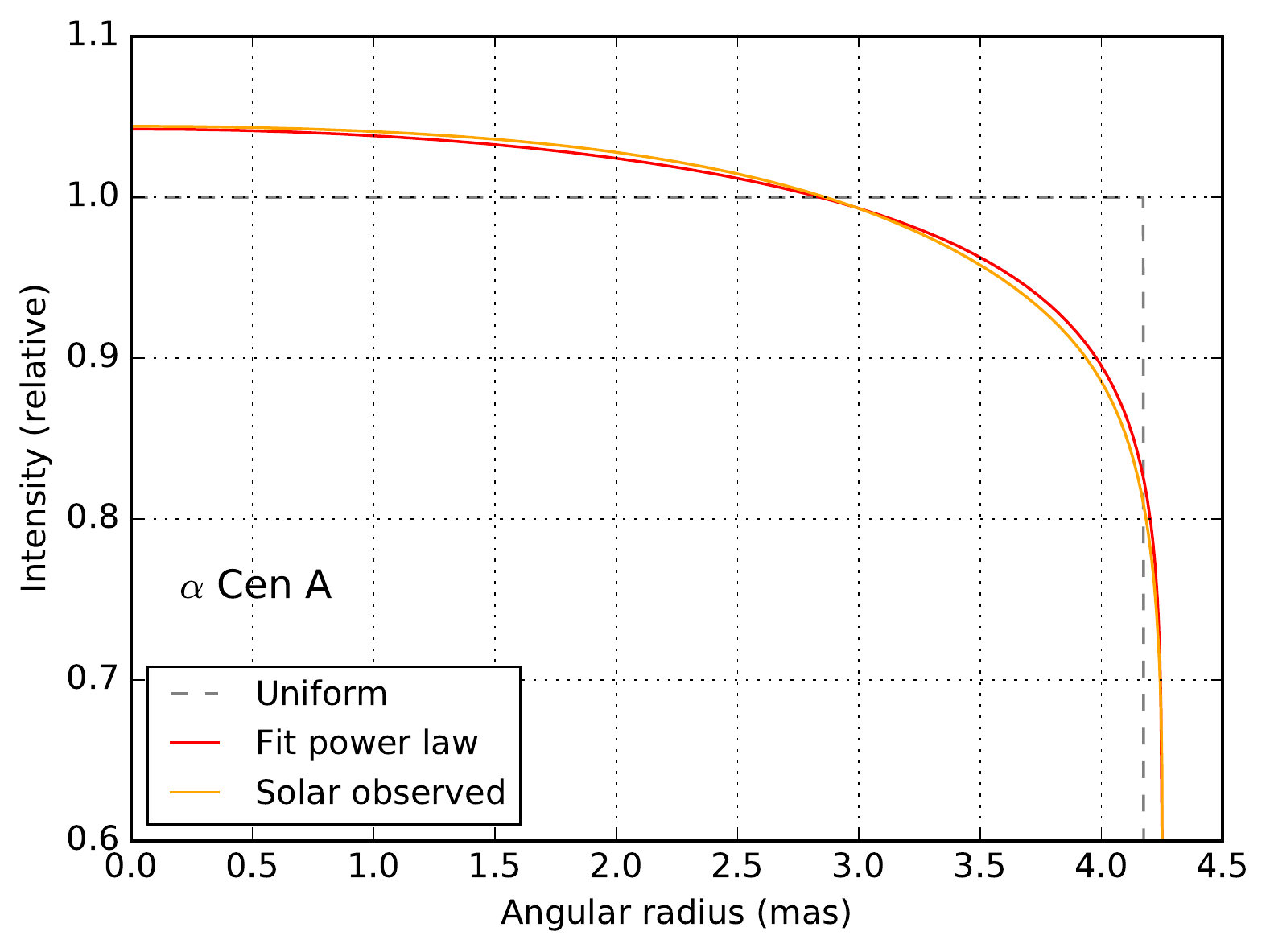}
        \includegraphics[width=\hsize]{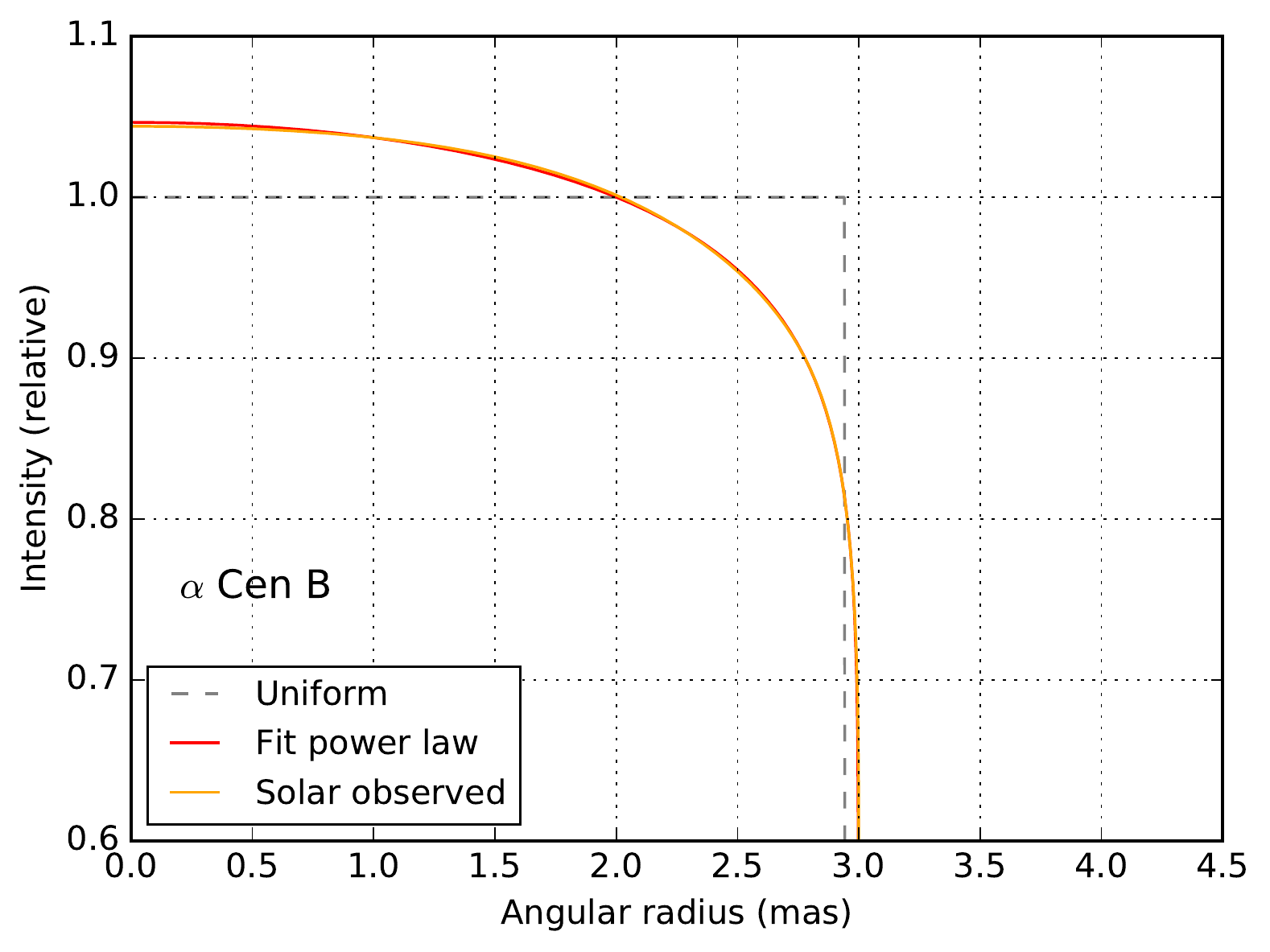}
        \caption{Comparison of the best-fit power law intensity profiles of $\alpha$\,Cen A and B (red curves) with the observed solar profile in the $H$ band (orange curves) measured by \citetads{1977SoPh...52..179P}.
        The horizontal scale is the same for both diagrams to show the difference in size of the two stars.
        \label{profiles}}
\end{figure}

\begin{table}
        \caption{Linear limb darkening parameters for $\alpha$\,Cen A and B from the literature.
        The measured values for $\alpha$\,Cen and the Sun (see Sect.~\ref{LDquality}) are listed in the last two lines.}
        \centering          
        \label{LDliterature}
        \begin{tabular}{lcc}
	\hline\hline
        \noalign{\smallskip}
        Ref. & Linear $u(A)$ & Linear $u(B)$ \\
         \hline
        \noalign{\smallskip}
         C95 & $0.310$ & $0.345$ \\
         C00 & $0.3532$ & $0.3922$ \\
         C11f & $0.2392$ & $0.2698$ \\
         C13f & $0.2545$ & $0.2839$ \\
         N13s & $0.4085$ & $0.4322$ \\
         N13p & $0.2336$ & $0.2736$ \\
         M15 & $0.3284$ & $0.3694$ \\
	R16 & $0.3555$ & $0.3888$ \\
         \hline
        \noalign{\smallskip}
         Measured & $0.1761 \pm 0.0062$ & $0.1907  \pm 0.0048$ \\
         \hline
        \noalign{\smallskip}
         Sun & \multicolumn{2}{c}{$0.2227 \pm 0.0005$} \\
         \hline
        \end{tabular}
	\tablebib{
	C00: \citetads{2000A&A...363.1081C};
	C11f: \citetads{2011A&A...529A..75C} (flux conservation);
	C13f: \citetads{2013A&A...552A..16C} (flux conservation);
	N13s: \citetads{2013A&A...556A..86N} (spherical);
	N13p: \citetads{2013A&A...556A..86N} (planar);
	M15: \citetads{2015A&A...573A..90M} (3D);
	R16: \citetads{2016MNRAS.456.1294R} (flux conservation).
	}
\end{table}

\subsection{Linear radii of $\alpha$\,Cen A and B}

The recently determined parallax of the $\alpha$\,Cen system by \citetads{2016kervella} of $\pi = 747.17 \pm 0.61$\,mas allows us to convert the measured LD angular diameters (using the power law LD) into linear radii.
We adopt the IAU convention \citepads{2016AJ....152...41P} for the nominal solar radius  ($\Rnom = {695\,700}$\,km), leading to the following conversion relation between the linear radius and the angular diameter:
\begin{equation}
R[R_\odot] = 9.3009345\ \frac{\theta[\mathrm{mas}]}{2}\ d[\mathrm{pc}].
\end{equation}
We obtain for $\alpha$\,Cen~A ($\pm \sigma_\mathrm{stat} \pm \sigma_\mathrm{syst}$),
\begin{eqnarray}
R_A & = & \radiusA \pm \radiusAerrstat \pm \radiusAerrsyst\ R_\odot
\end{eqnarray}
and for $\alpha$\,Cen~B,
\begin{eqnarray}
R_B & = & \radiusB \pm \radiusBerrstat \pm \radiusBerrsyst\ R_\odot
\end{eqnarray}
The error bars are dominated by the systematic uncertainty on the effective wavelength of PIONIER.
\citetads{2003A&A...404.1087K} determined $R_A = 1.224 \pm 0.003\ R_\odot$, $R_B = 0.863 \pm 0.005\ R_\odot$ (\citeads{2006A&A...446..635B} found $R_B = 0.863 \pm 0.003\ R_\odot$) from VLTI/VINCI measurements in the near-infrared $K$ band ($\lambda = 2.2\,\mu$m), assuming the same parallax as in the present work.
These values are remarkably identical to the present measurements (offsets of $+0.10\sigma$ and $-0.04\sigma$, respectively), indirectly confirming the quality of our wavelength calibration (Sect.~\ref{wavecalibration}).

The ratio of the radii of $\alpha$\,Cen A and B is an interesting differential quantity as it is insensitive to the wavelength calibration of the instrument and the parallax,
\begin{equation}
\frac{R_A}{R_B} = \ratioR \pm \ratioRerr.
\end{equation}
The accuracy of this ratio ($\ratioRrelerr\%$) is limited by the statistical uncertainties that are very small in our case.
This ratio is in perfect agreement with the measurement by \citetads{2003A&A...404.1087K} ($R_A/R_B = 1.418 \pm 0.009$) and within $1.2\sigma$ of the prediction $R_A/R_B = 1.435 \pm 0.014$ by \citetads{2002A&A...392L...9T}.
This quantity is well suited to constrain the models of the $\alpha$\,Cen pair, as both stars have the same age and the same initial composition. Therefore, their evolution can easily be traced in parallel using numerical models, where the only difference in the input parameters are their initial masses.

\subsection{Luminosities and effective temperatures}

We derive the effective temperatures of $\alpha$\,Cen~A and B considering the bolometric flux values determined by \citetads{2013ApJ...771...40B} and adopted by \citetads{2015A&A...582A..49H}: $F_\mathrm{bol}[A] = (27.16 \pm 0.27) \times 10^{-9}$\,W\ m$^{-2}$ and $F_\mathrm{bol}[B] = (8.98 \pm 0.12) \times 10^{-9}$\,W\ m$^{-2}$.
The parallax $\pi = 747.17 \pm 0.61$\,mas is taken from \citetads{2016kervella} giving luminosities of
\begin{eqnarray}
L_A & = & \lumA \pm \lumAerr\,L_\odot \\
L_B & = & \lumB \pm \lumBerr\,L_\odot
\end{eqnarray}
These values assume a nominal solar luminosity $\Lnom = 3.828 \times 10^{26}$\,W.
A straight application of the Stefan-Boltzmann law gives effective temperatures of
\begin{eqnarray}
T_{\mathrm{eff}}[A] & = & \teffA \pm \teffAerr\ \mathrm{K} \\
T_{\mathrm{eff}}[B] & = & \teffB \pm \teffBerr\ \mathrm{K}
\end{eqnarray}
in perfect agreement with \citetads{2015A&A...582A..49H}.


\section{Conclusions}

\begin{table}
        \caption{Fundamental parameters of $\alpha$\,Cen A and B.}
        \centering          
        \label{paramsAB}
        \begin{tabular}{lll}
	\hline\hline
        \noalign{\smallskip}
        Parameter & Value & Ref. \\
         \hline
        \noalign{\smallskip}
	        \multicolumn{3}{c}{$\alpha$\,Cen AB} \\
        \noalign{\smallskip}
	Parallax & $717.17 \pm 0.61$\,mas & K16a \\
        \noalign{\smallskip}
	Distance & $1.3384 \pm 0.0011$\,pc & K16a \\
	 & $4.1298 \pm 0.0034 \times 10^{16}$\,m & K16a \\
        \noalign{\smallskip}
         \hline
        \noalign{\smallskip}
	        \multicolumn{3}{c}{$\alpha$\,Cen A} \\
        \noalign{\smallskip}
         Mass & $\massA \pm \massAerr\ M_\odot$ & K16a \\
            & $\massAkg \pm \massAerrkg \times 10^{30}$\,kg &  \\
         \noalign{\smallskip}
        Radius & $\radiusA \pm \radiusAerrtot\ R_\odot$ & K16a, K16b \\
            & $\radiusAm \pm \radiusAerrm \times 10^{8}$\,m &  \\
        \noalign{\smallskip}
         Luminosity & $\lumA \pm \lumAerr\ L_\odot$ & B13, K16a \\
            & $\lumAW \pm \lumAerrW \times 10^{26}$\,W & \\
        \noalign{\smallskip}
         $T_\mathrm{eff}$ & $\teffA \pm \teffAerr$\,K & B13, K16b \\
        \noalign{\smallskip}
         $\log g$ [cgs] & $\loggA \pm \loggAerr$ & K16a, K16b \\
        \noalign{\smallskip}
         \hline
        \noalign{\smallskip}
	        \multicolumn{3}{c}{$\alpha$\,Cen B} \\
        \noalign{\smallskip}
         Mass & $\massB \pm \massBerr\ M_\odot$ & K16a \\
            & $\massBkg \pm \massBerrkg \times 10^{30}$\,kg &  \\
        \noalign{\smallskip}
         Radius & $\radiusB \pm \radiusBerrtot\ R_\odot$ & K16a, K16b \\
            & $\radiusBm \pm \radiusBerrm \times 10^{8}$\,m &  \\
        \noalign{\smallskip}
         Luminosity & $\lumB \pm \lumBerr\ L_\odot$ & B13, K16a \\
            & $\lumBW \pm \lumBerrW \times 10^{26}$\,W & \\
        \noalign{\smallskip}
         $T_\mathrm{eff}$ & $\teffB \pm \teffBerr$\,K & B13, K16b \\
        \noalign{\smallskip}
         $\log g$ [cgs] & $\loggB \pm \loggBerr$ & K16a, K16b \\
        \noalign{\smallskip}
         \hline
        \end{tabular}
	\tablebib{
	B13: \citetads{2013ApJ...771...40B},
	K16a: \citetads{2016kervella},
	K16b: present work.
	}
\end{table}

We presented new high-accuracy interferometric measurements of the angular diameters and limb darkening parameters of $\alpha$\,Centauri A and B in the infrared $H$ band.
The accuracy on the angular diameters ($0.4\%$) is presently limited by the wavelength calibration of the PIONIER instrument, but it will be significantly improved when the parallax of the dimensional calibrator \object{HD\,123999} will be available from \emph{Gaia} \citepads{2016arXiv160904153G}.
The VLTI/GRAVITY beam combiner \citepads{2011Msngr.143...16E} will also soon overcome this limitation in the infrared $K$ band ($\lambda = 2.2\,\mu$m) thanks to its highly accurate laser-referenced wavelength calibration ($\sigma<0.1\%$).

We observe a significant discrepancy of the measured linear LD parameters $u$ with respect to model predictions from the literature, which systematically overestimate the limb darkening of $\alpha$\,Cen~A and B.
Setting the value of $u$ from existing tabulated model atmospheres results in an overestimation of the LD angular diameter by 0.5\% compared to the more realistic power law profile.
Over the complete sample of LD angular diameter values listed in Table~\ref{LDmodels} for $\alpha$\,Cen A and B (considering all parametric models), we observe an amplitude of 1\% between the extreme values.
This difference is small and gives confidence in the existing angular diameter measurements of the literature at this level of accuracy.
However, the shape of the intensity profile itself, reflected on the interferometric visibility function, is incorrectly predicted by the models.
This difference may be partly due to the mathematical technique used to extract, for instance, the linear LD parameter $u$ from the stellar atmosphere models.
But higher order approximations (e.g.~four-parameter) also fail to reproduce the observed visibilities satisfactorily, particularly the second lobe of the visibility function.
This implies that the underlying atmosphere models deviate from the real intensity profiles of $\alpha$\,Cen, and we note that similar discrepancies are observed on the Sun.
The observed discrepancies indicate that the predictive accuracy of the current generation of model atmospheres may be significantly lower than expected.
This is likely to be more critically the case for stars with parameters that are very different from those of the Sun (e.g., cooler stars with molecular envelopes) and for wavelength regions more complex to model than the near-infrared (e.g., the ultraviolet and visible).
The high-precision modeling of exoplanet transits \citepads{2015MNRAS.450.1879E} and eclipsing binaries \citepads{2013Natur.495...76P,2016arXiv160906645G} or the calibration of surface-brightness color relations \citepads{2004A&A...426..297K,2008A&A...491..855K, 2013ApJ...771...40B} will require increasingly accurate LD models.
Stellar parallaxes more accurate than 1\% are still relatively rare, but thanks to \emph{Gaia} they will soon become much more common, and a higher accuracy on the LD will be a requirement for unbiased stellar population studies.
The measured PIONIER squared visibilities provide very valuable benchmarks to validate future evolutions of atmosphere models of late-type stars.
We note that the existing solar LD measurements between wavelengths of 1 to $4\,\mu$m are now relatively old and would certainly benefit from new observations using modern detectors.

The measured photospheric radii of $\alpha$\,Cen~A and B are however in perfect agreement with the values obtained 13\,years ago by \citetads{2003A&A...404.1087K} in the $K$ band.
These new measurements are independent of model atmospheres for the prediction of LD correction coefficients as they are now measured.
Together with the parallax and masses recently reported by \citetads{2016kervella}, as well as spectroscopic studies, the determined radii complete the calibration of the fundamental parameters of both components of $\alpha$\,Centauri (Table~\ref{paramsAB}).

\begin{acknowledgements}
Based on observations collected at the European Organisation for Astronomical Research in the Southern Hemisphere under ESO programs 096.D-0299(A), 097.D-0350(A) 097.D-0350(B) and 097.D-0350(C).
We thank the ESO Paranal team for their excellent support during the observations of $\alpha$\,Cen with PIONIER and, in particular, Thomas Rivinius.
We are grateful to Antoine M\'erand for providing the orbital prediction routines used to analyze the observations of HD\,123999.
This research has made use of the Jean-Marie Mariotti Center \texttt{LITpro} service co-developed by CRAL, IPAG and LAGRANGE.
This research has made use of the Jean-Marie Mariotti Center \texttt{Aspro} service \footnote{Available at \url{http://www.jmmc.fr/aspro}}.
We acknowledge financial support from the ``Programme National de Physique Stellaire'' (PNPS) of CNRS/INSU, France.
This research made use of the SIMBAD and VIZIER databases (CDS, Strasbourg, France) and NASA's Astrophysics Data System.
\end{acknowledgements}

%
\bibliographystyle{aa} 
\bibliography{BiblioAlfCen.bib} 
%

\begin{appendix}
\section{PIONIER astrometry of HD\,123999}

The PIONIER astrometry of HD\,123999\,B relative to A is presented in Table~\ref{12boo-positions}.

\begin{table*}
        \caption{Positions of HD\,123999\,B relative to A measured with PIONIER. These positions are not corrected for the $\gamma$ scaling coefficient.}
        \centering          
        \label{12boo-positions}
        \begin{tabular}{lccc}
	\hline\hline
        \noalign{\smallskip}
        MJD-57\,000$^a$ & dRA [mas] & dDec [mas] & $f(A)/f_\mathrm{tot}$(H)\ $^b$ \\
         \hline         
        \noalign{\smallskip}
439.346279 & $+3.5132 \pm 0.0083$ & $+0.1105 \pm 0.0110$ & $0.6160 \pm 0.0043$ \\
439.363306 & $+3.3658 \pm 0.0410$ & $+0.0173 \pm 0.0585$ & $-$ \\
448.335568 & $+3.1083 \pm 0.0041$ & $-0.2885 \pm 0.0080$ & $0.6059 \pm 0.0029$ \\
448.362032 & $+3.1360 \pm 0.0047$ & $-0.2967 \pm 0.0059$ & $0.6071 \pm 0.0018$ \\
449.341644 & $+3.5605 \pm 0.0063$ & $+0.3777 \pm 0.0124$ & $0.6222 \pm 0.0047$ \\
449.366629 & $+3.5683 \pm 0.0049$ & $+0.3841 \pm 0.0114$ & $0.5885 \pm 0.0050$ \\
539.056380 & $-2.3955 \pm 0.0017$ & $+0.1811 \pm 0.0046$ & $0.6152 \pm 0.0012$ \\
539.068678 & $-2.4155 \pm 0.0024$ & $+0.1490 \pm 0.0076$ & $0.6106 \pm 0.0023$ \\
539.079484 & $-2.4334 \pm 0.0073$ & $+0.1490 \pm 0.0128$ & $0.6083 \pm 0.0024$ \\
539.089021 & $-2.4451 \pm 0.0061$ & $+0.1144 \pm 0.0130$ & $0.6019 \pm 0.0029$ \\
539.098338 & $-2.4675 \pm 0.0040$ & $+0.0952 \pm 0.0105$ & $0.6145 \pm 0.0019$ \\
 \hline
        \end{tabular}
        \tablefoot{
        $^a$ MJD is the average modified julian date.
	$^b$ $f(A)/f_\mathrm{tot}$ is the flux from component A relative to the total flux.
        }
\end{table*}
\end{appendix}

\end{document}